\makeatletter
\declare@file@substitution{revtex4-1.cls}{revtex4-2.cls}
\makeatother
\documentclass[twocolumn, times, tighten,twocolappendix]{aastex63}
\usepackage{amsmath}
\usepackage{amssymb,txfonts}
\usepackage[T1]{fontenc}
\usepackage{graphicx,multirow,hyperref,url,color,xspace}
\usepackage{subfigure}
\usepackage{overpic}
\DeclareMathAlphabet      {\mathbfit}{OML}{cmm}{b}{it}
\usepackage{lineno}

\newcommand{\g}{$\gamma$}
\newcommand{\msun}{{M}_{\sun}}

\newcommand{\fermi}{{\textit{Fermi}}\xspace}
\newcommand{\agile}{{\textit{AGILE}}\xspace}

\defcitealias{Zdziarski18}{Z18}
\newcommand{\aaz}{{\citetalias{Zdziarski18}}\xspace}
\defcitealias{Antokhin22}{A22}
\newcommand{\ant}{{\citetalias{Antokhin22}}\xspace}

\begin{document}

\title{Two models for the orbital modulation of $\gamma$-rays in Cyg X-3}

\author[0000-0003-0102-5579]{Anton Dmytriiev}
\affiliation{Centre for Space Research, North-West University, Potchefstroom, 2520, South Africa; \href{mailto:anton.dmytriiev@nwu.ac.za}{anton.dmytriiev@nwu.ac.za}}

\author[0000-0002-0333-2452]{Andrzej A. Zdziarski}
\affiliation{Nicolaus Copernicus Astronomical Center, Polish Academy of Sciences, Bartycka 18, PL-00-716 Warszawa, Poland; \href{mailto:aaz@camk.edu.pl}{aaz@camk.edu.pl}}

\author[0000-0001-9689-2194]{Denys Malyshev}
\affiliation{Institut f\"{u}r Astronomie und Astrophysik T\"{u}bingen, Universit\"{a}t T\"{u}bingen, Sand 1, D-72076 T\"{u}bingen, Germany}

\author[0000-0002-6043-5079]{Valentí Bosch-Ramon}
\affiliation{Departament de Física Quàntica i Astrofísica, Institut de Ciències del Cosmos, Universitat de Barcelona, Martí i Franquès 1, 08028 Barcelona, Spain}
\affiliation{Instituto Argentino de Radioastronomía, CONICET, Berazategui Partido, Buenos Aires Province, Argentina}

\author[0000-0002-9735-3608]{Maria Chernyakova}
\affiliation{School of Physical Sciences and Centre for Astrophysics \& Relativity, Dublin City University, Glasnevin, D09 W6Y4, Ireland}
\affiliation{Dublin Institute for Advanced Studies, 31 Fitzwilliam Place, Dublin 2, Ireland}

\begin{abstract}
We model the currently available $\gamma$-ray data from the Fermi Large Area Telescope on Cyg X-3. Thanks to its very strong $\gamma$-ray activity during 2018--2021, the data quality has significantly improved. We study the strong orbital modulation of the $\gamma$-rays observed during at high $\gamma$-ray fluxes. The modulation, as found earlier, is well modeled by anisotropic Compton scattering of the donor blackbody emission by relativistic electrons in a jet strongly misaligned with respect to the orbital axis. We confirm that this model fits well both the average $\gamma$-ray modulation light curve and the spectrum. However, we find that if the jet is aligned with the spin axis of a rotating black hole, it would undergo geodetic precession with the period of $\sim$50 years. However, its presence is ruled out by both the $\gamma$-ray and radio data. Therefore, we consider an alternative model in which the average jet direction jet is aligned, but it is bent to outside the orbit owing to the thrust of the donor stellar wind, and thus precesses at the orbital period. The $\gamma$-ray modulation appears then owing to the variable Doppler boosting of synchrotron self-Compton jet emission. The model also fits well the data. However, the fitted bending angle is much larger than the theoretical one based on the binary and wind parameters as currently known. Thus, both models disagree with important aspects of our current theoretical understanding of the system. We discuss possible ways to find the correct model.
\end{abstract}

\section{Introduction} \label{Intro}

The high-mass binary Cyg X-3 is one of the first discovered \citep{Giacconi67} X-ray binaries. Still, after many years of intense research, it remains to be a puzzling (and unique) system. The nature of its compact object remains uncertain. It can be either a neutron star or a black hole (BH), see \citet{Koljonen17} and \citet{Antokhin22}, hereafter \ant, for discussions. Still, those authors favored the presence of a BH, which is also supported by various aspects of the X-ray and radio emission \citep{Hj08,Hj09,SZ08,SZM08,Koljonen10}. Its donor is a Wolf-Rayet (WR) star \citep{vanKerkwijk92, vanKerkwijk96, Koljonen17}, which makes it the only known binary system in the Milky Way consisting of a WR star and a compact object, and a candidate for a future merger \citep{Belczynski13}. The compact object accretes a fraction of the stellar wind from the donor. \ant estimated the total mass of the system as $18.8\msun$, and the compact object mass as $M_{\rm c}\approx 7.2\msun$, implying it is a BH. However, they also considered lower masses possible if the wind is significantly clumped. Its distance has recently been determined based on a radio parallax as $D\approx 9.7\pm 0.5$ kpc and based on Galactic proper motions and line-of-sight radial velocity measurements as $9 \pm 1$ kpc \citep{Reid23}. We hereafter assume $D=9$\,kpc, $M_*=11.6\msun$, $M_{\rm c}=7.2\msun$.  

Cyg X-3 is also the brightest and most variable radio source among X-ray binaries, showing resolved relativistic jets \citep{Mioduszewski01, Miller-Jones04}. Its jets also emit high-energy \g-rays, as first discovered by \citet{Mori97}, and later confirmed by observations with the \fermi Large Area Telescope (LAT; \citealt{Fermi09}), and \agile \citep{Tavani09}. As found by \citet{Fermi09}, the \g-rays emitted during flaring epochs are strongly orbitally modulated. The modulated emission was explained by \citet{Dubus10} as anisotropic Compton scattering of blackbody photons from the donor by relativistic electrons with a power-law distribution accelerated in the jet (which we hereafter refer to as either the blackbody Compton model or Model 1). The peak of the emission of a jet aligned with the orbital axis would be at the superior conjunction of the compact object. However, \citet{Dubus10} found that the peak was clearly before the superior conjunction, indicating that the jet is inclined with respect to the orbital axis. With the number of high-energy \g-rays increasing with time, the statistical accuracy of the orbital modulation increased, and this model was fitted to the updated data by \citet{Zdziarski18}, hereafter \aaz. 

\begin{figure}[t!]
\centerline{
\includegraphics[width=\columnwidth]{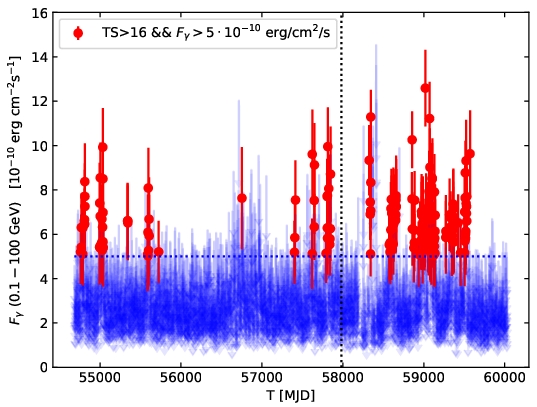}}
  \caption{The LAT high-energy \g-ray light curve of Cyg X-3 from the beginning of the \fermi observations until MJD 60200. The red and blue symbols represent the detections within a day and upper limits, respectively. Our criterion defining the \g-ray bright state for 1-d integration is shown by the horizontal dotted line. \aaz analyzed the data until MJD 57982, which is marked by the vertical dotted line. We see that the later observations have multiplied the number of detected photons by a substantial factor. 
}\label{lc}
\end{figure}

The observations analyzed by \aaz were until MJD 57982. As we have found out by analyzing the LAT data, Cyg X-3 has shown very strong activity after that date, see Figure \ref{lc}. The new data allow us to increase the number of detected \g-ray photons by a large factor, and, consequently, to strongly reduce the errors on the light curve averaged and folded over the orbital period.

In our modeling, we consider an important consequence of the jet--binary axis misalignment. Jets powered by the rotation of BHs \citep{BZ77} are launched along the BH spin axis \citep{McKinney13}. Jets launched via poloidal magnetic field of accretion disks \citep{BP82} will initially be directed along the inner disk rotation axis, which could, in turn, be also aligned with the BH spin \citep{Bardeen75}. In those cases, the observed jet--orbit misalignment would correspond to the BH spin--orbit one. However, spin precession due to the binary angular momentum--compact object spin coupling has to take place in general relativity, and the period of such geodetic precession for Cyg X-3 is as short as $\sim$50 years. Such precession is clearly not seen in the \fermi data, which span $\approx$14 years, i.e., more than a quarter of that period. It is also not seen in the $\sim$30 years of the radio imaging data, in which the jet position angle remains constant along the north-to-south direction. As shown in this work, these observations present an argument against the misaligned jet model. 

Therefore, we also consider an alternative model of the orbital modulation in which it is due to the Doppler boosting of the jet synchrotron self-Compton (SSC) emission. This could be achieved if the jet is bent to outside by the thrust of the stellar wind \citep{Yoon15, Yoon16, Bosch-Ramon16, Molina18, Molina19, Barkov22}, with the wind being very strong in this binary. In addition, the Coriolis force due to the binary rotation bends the jet opposite to the rotation direction. We hereafter refer to it as either the orbital precession model or Model 2. We have found that the resulting modulated Doppler boosting can fit the data at an accuracy similar to the blackbody Compton model. In this model, the average jet direction is fully aligned with the orbital axis, and thus no geodetic precession is expected. An important prediction of this model is an orbital modulation of the associated synchrotron emission, with the orbital light curve similar to that of the $\gamma$-rays. However, we find that the fitted bending angle is much larger than that expected theoretically based on the likely wind and jet parameters. This, in turn, represents a strong argument against this model. 

We present our detailed results for both models. We then discuss possible solutions of their problems for each of them. According to our present knowledge, we cannot determine which model is more likely.

\begin{figure}[t!]
\centerline{
\includegraphics[width=\columnwidth]{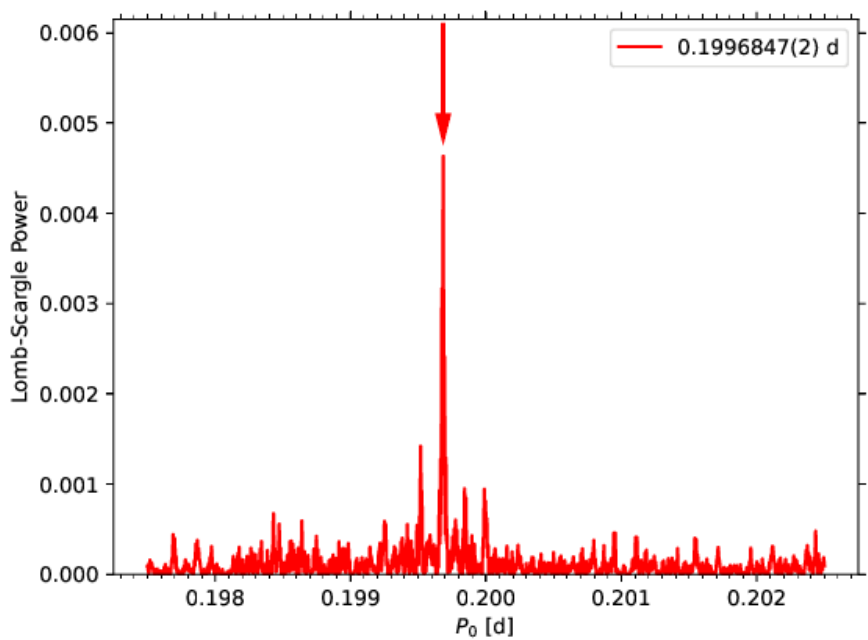}}
  \caption{The Lomb-Scargle periodogram in the \g-ray bright state in the 0.1--100 GeV range, calculated accounting for the secular orbital period increase and taking into account the measurement uncertainties. The only significant peak is equal within the fitted uncertainties to the orbital period of Cyg X-3.  
}\label{LS}
\end{figure}

\section{The data analysis} \label{Data}

We analyze the data from the direction of Cyg X-3 for MJD 57982--60201, see Figure \ref{lc}. We used the Fermi Science Tools v. 2.2.0. Details of the analysis generally follow those given in \aaz. For the spectral analysis, we used the energy range of 0.05--500\,GeV and for the timing analysis, that of 0.1--100\,GeV. In this paper, we analyze only \g-ray bright states, which we define in the same way as the `flaring state' in \aaz. The minimum level of this state is shown by the horizontal line in Figure \ref{lc}. These fluxes correspond to $F(0.1$--100\,GeV$) > 5\times 10^{-10}$\,erg\,cm$^{-2}$\,s$^{-1}$ and the test statistics \citep{Mattox96} of ${\rm TS}>16$, i.e., the significance $>4\sigma$. We obtain the average spectrum for that state with positive detections in the 0.05--20\,GeV range, and upper limits at higher energies. 

We used the quadratic ephemeris given by model 4 of \citet{Antokhin19} and searched for periodicity in the LAT light curve taking into account their rate of the period increase. We used the Lomb-Scargle method. Our results are shown in Figure \ref{LS}. We have found the period of $P_0= 0.1996847(2)$\,d, which is in full agreement with that of \citet{Antokhin19}, $P_0= 0.199684622(15)$\,d. We also tested for the effect of using their quadratic+sinusoidal ephemeris, but we found its effect to be very minor. Then, we calculated the folded and averaged light curve in 10 phase bins.

\section{Theoretical description} \label{description}

\begin{figure}[t!]
\centerline{
\includegraphics[width=\columnwidth]{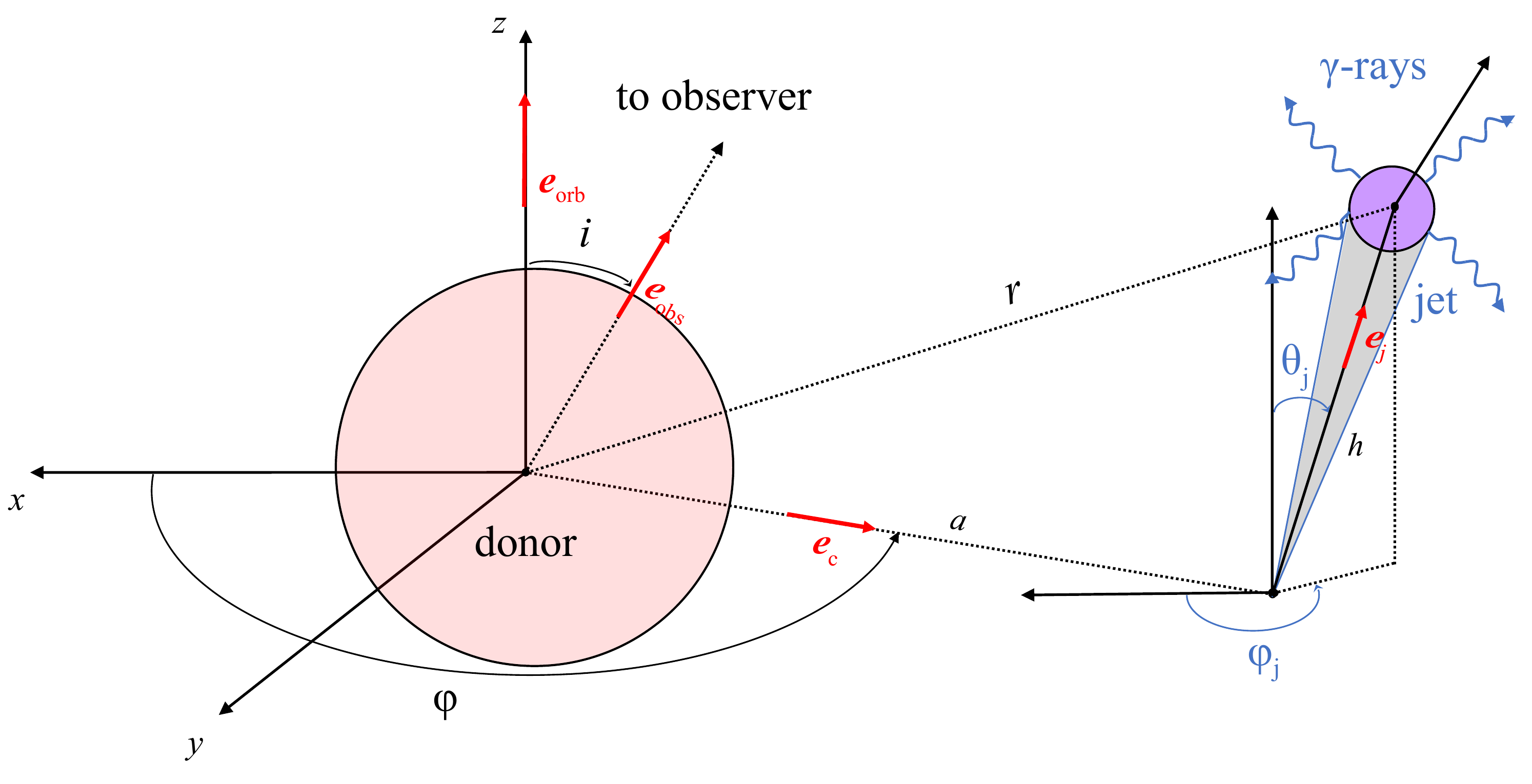}}
  \caption{The geometry corresponding to emission in a compact region of the jet (the counterjet is not shown) for the assumed counter-clockwise rotation. The axes $x$ and $y$ are in the binary plane, and the $+z$ direction is parallel to the orbital axis. The $-x$ direction gives the projection of the direction toward the observer onto the binary plane. The observer is at an angle, $i$, with respect to the orbital axis, $\varphi$ is the orbital phase (the superior conjunction is at $\varphi=0$, and the shown configuration is close to the inferior conjunction), $\theta_{\rm j}$ and $\varphi_{\rm j}$ are the polar and azimuthal angles of the jet with respect to the orbital axis, $a$ is the binary separation, $h$ is the distance of the \g-ray source from the center of the compact object, ${\mathbfit e}_{\rm orb}$ is a unit vector perpendicular to the orbital plane, ${\mathbfit e}_{\rm obs}$, ${\mathbfit e}_{\rm c}$ and ${\mathbfit e}_*$ point from the donor toward the observer, the center of the compact object, and the \g-ray source, respectively, and ${\mathbfit e}_{\rm j}$ points from the center of the compact object toward the \g-ray source.  
}\label{geo}
\end{figure}

\subsection{The geometry}
\label{geometry}

Our assumed geometry is shown in Figure \ref{geo}. The binary rotates with the period $P$, and the azimuth of the rotation with respect to the superior conjunction is $\varphi$. We assume no eccentricity and the rotation to be counter-clockwise, as found by \citet{Veledina24}. The \g-ray emitting region in the jet is at the distance $h$ from the compact object, which itself is at the binary separation, $a$, from the center of the donor. The jet is inclined with respect to the orbital axis as given by the polar and azimuthal angles, $\theta_{\rm j}$ and $\varphi_{\rm j}$, respectively, and the distance of the emitting region from the center of the donor is $r$. The jet polar angle, $\varphi_{\rm j}$, is either fixed in the case of the blackbody Compton model or dependent on the orbital phase in the orbital precession model. 

The unit vectors pointing towards the observer, from the stellar center to the compact object, along the jet, and along $z$ (parallel to the orbital axis), are
\begin{align}
    & {\mathbfit e}_{\rm obs} = (-\sin i, 0, \cos i), \quad {\mathbfit e}_{\rm c} = (\cos \varphi, \sin \varphi, 0), \\
    & {\mathbfit e}_{\rm j} = (\sin \theta_{\rm j} \cos \varphi_{\rm j},  \sin \theta_{\rm j} \sin \varphi_{\rm j}, \cos \theta_{\rm j}),\quad {\mathbfit e}_{\rm orb}=(0,0,1),
\end{align}
respectively. We make a simplifying assumption that the modulated emission at a given frequency originates at a single distance, $h$, from the BH. The vector connecting the donor center with the emission point is $a {\mathbfit e}_{\rm c} + h {\mathbfit e}_{\rm j}$. The length of this vector gives the distance of that point from the donor, and its square is given by
\begin{equation}
    r^2 = h^2 + a^2 + 2 h a \sin \theta_{\rm j} \cos(\varphi_{\rm j} - \varphi).
\end{equation}

An important issue concerns the determination of the phase of the superior conjunction. All of the determinations of the ephemeris so far have been based on fitting average profiles of the X-ray emission in various bands (see \citealt{Bhargava17, Antokhin19} for the most recent work). The X-ray modulation appears due to the bound-free absorption and scattering by the stellar wind emitted by the donor. Since those profiles are not symmetric, a template designed by \citet{vanderKlis89} was used for the fits. However, the phases corresponding to the observed minimum fluxes using that template are {\it not\/} null (see, e.g., the profiles observed in various X-ray bands and the hard and soft X-ray states of the source in \citealt{Zdziarski12b}). This issue was studied in detail by \ant. They used the X-ray data from the All Sky Monitor \citep{ASM} onboard {\it Rossi X-ray Timing Explorer}, and the infrared data acquired by them. They estimated the superior conjunction based on their three-component wind model fitted to those data to be $\phi_0/2 \pi=-0.066\pm 0.006$, where $\phi$ is the orbital phase defined by the ephemeris. Since the highest \g-ray flux in our folded/averaged light curve is in the phase bin with $\phi/2\pi=-0.15\pm 0.05$, the correction to the phase of the superior conjunction is significant. Thus, the true orbital phase of the binary is given by  
\begin{equation}
\varphi/2\pi=\phi/2\pi +0.066\pm 0.006.
\label{superior}
\end{equation} 

When the jet is inclined with respect to the orbital axis, its viewing angle is different from that of the orbital axis itself. It equals to
\begin{equation}
i_{\rm j}=\arccos (\cos i \cos\theta_{\rm j} - \cos\varphi_{\rm j} \sin i \sin\theta_{\rm j}).\label{ijet}
\end{equation}
Furthermore, the position angle on the sky of the binary differs from that of the jet. The difference between the jet and the binary position angles, $\Delta\lambda$, is given by
\begin{equation}
\cos \Delta\lambda= \frac{\cos i\cos \varphi_{\rm j}  \sin \theta_{\rm j}+\sin i \cos \theta_{\rm j}}{\sqrt{1-\left(\cos i \cos \theta_{\rm j}-\sin i\cos \varphi_{\rm j} \sin \theta_{\rm j}\right)^2}}.
\label{delta}
\end{equation}

The value of the binary separation follows via the Kepler law from the assumed total mass of $18.8\msun$, yielding $a\approx 2.66\times 10^{11}$\,cm. We either assume the binary inclination of $i=30\degr$ (\ant) or treat it as a free parameter.

\subsection{The electrons}
\label{electrons}

The electrons in the emitting regions are assumed to be isotropic in the comoving frame, and move with bulk the velocity $\beta_{\rm j} c$. Also, they are assumed to have a power law distribution, $N(\gamma)=K\gamma^{-p}$ from $\gamma_{\rm min}$ to $\gamma_{\rm max}$. Hereafter, $N(\gamma)$ refers to all of the power-law electrons in the source. However, we normalize the electron distribution to the kinetic energy content (i.e.\ excluding the rest energy) between $\gamma = 1$ to $\gamma_{\rm max}$, $E_{\rm e}$, instead of $K$. 

We also take into account the electrons cooled by the blackbody Compton, synchrotron, SSC and adiabatic losses to below $\gamma_{\rm min}$. This is important since these particles contribute significantly to the emission at low energies. The electron distribution below $\gamma_{\rm min}$ is determined by a kinetic equation, where the radiative and adiabatic losses compete. Solving the kinetic equation with no injection of fresh electrons below $\gamma_{\rm min}$ yields approximately a broken power-law distribution, with $N_{\rm e} \propto \gamma^{-2}$ in the range $\gamma_{\rm br} \leq \gamma < \gamma_{\rm min}$ (where the radiative losses dominate), and $N_{\rm e} \propto \gamma^{-1}$ in the range $1 \lesssim \gamma < \gamma_{\rm br}$ (where the adiabatic losses dominate). The former follows from both the Compton losses in the Thomson regime and the synchrotron ones having $\dot \gamma \propto -\gamma^2$. The position of the cooling break $\gamma_{\rm br}$ is determined self-consistently in each simulation by equating the timescales of total radiative losses and adiabatic losses. We verify that the Compton scattering around $\gamma_{\rm br}$ is in the Thomson regime. 

\subsection{Radiation}
\label{radiation}

We assume a one-zone \g-ray emitting region in our model. To validate this assumption, we examined the importance of effects of propagation of electrons and the evolution of their distribution along the considered source \citep{Khangulyan18}, which, if significant, might lead to a departure from the one-zone model. We find that these propagation effects have only a minor impact in our specific case. This is due to the fact that the jet velocity $\beta_{\rm j}$ in all our fits remains below the value identified in \citet{Khangulyan18}, above which such effects become non-negligible.

The treatment of the anisotropic Compton scattering off stellar blackbody photons follows that of \citet{Zdziarski12a}, which employs the full Klein-Nishina (KN) cross section integrated over a broken power law electron distribution. The blackbody photons are calculated for a donor with radius $R_*$ and temperature $T_*$, but we approximate that emission as coming from a point source. For that process, the total number of electrons determines the observed flux. The emission region has to have a low vertical extent since the orbital profile of the blackbody Compton emission depends sensitively on the height.

For an assumed value of the magnetic field strength, $B$, we compute synchrotron and SSC emission assuming that the jet emitting region has a cylindrical geometry (which is a local approximation to a section of a cone) and is filled with a tangled and uniform magnetic field. The electrons can also be present outside that region, but we neglect their presence in our treatment. The emission is therefore isotropic in the comoving frame. The radius of the cylinder is defined by the opening angle of the jet $\alpha_{\rm j}$ as $R = \alpha_{\rm j} h$, and for the height of the cylinder we arbitrarily choose $\Delta h = (1/3)h$. We note that the observed synchrotron spectrum for given $B$ and $K$ does not depend on neither $\Delta h$ nor $h$, with its luminosity being $L_{\rm S}\propto B^2 K$, apart from the synchrotron self-absorbed domain at low frequencies. Then, the shape of the SSC spectrum is also independent of them. However, the ratio of the two luminosities (the Compton dominance), is
\begin{equation}
\frac{L_{\rm SSC}}{L_{\rm S}}\propto\frac{L_{\rm S}}{\alpha_{\rm j} h\Delta h B^2}\propto\frac{K}{\alpha_{\rm j} h\Delta h}.
\label{ratio}
\end{equation}
In our Model 2, with precessing jet, $L_{\rm SSC}$ is fixed by the observed \g-ray flux, and hence the predicted $L_{\rm S}$ increases when increasing $\alpha_{\rm j} h\Delta h$.

For the single electron synchrotron emissivity, we employ an approximation from eq.~(3.40) in \citet{Boettcherbookjets2012}. We include the synchrotron self-absorption effect by using eq.~(3.42) in \citet{Boettcherbookjets2012}. To obtain the synchrotron radiation energy density, we use the solution of the radiation transfer equation for cylindrical geometry \citep[e.g., eq.~14 in][]{Chiaberge1999}, while to calculate the observed synchrotron flux in the absorbed regime, we follow eqs.~(18--19) of \citet{ZLS12}. The SSC computation takes into account the KN cross section and follows eqs.\ (8--12) in \citet{Katarzynski01} (though excluding the scaling factor of 3/4 for spherical geometry present in their eq.~11). The emission is transformed to the observer's frame, including contributions from both the jet and counterjet. 

We also check the importance of internal \g--\g absorption of high-energy \g-rays on the soft synchrotron photon field. We find that this effect is negligible within our studied energy range, and becomes pronounced only above 10--20 GeV. At those energies, however, the the source is no longer detected.

\subsection{Additional observational constraints}
\label{observational}

Apart from the LAT \g-ray data, we impose some observational constraints in other spectral regions. First, the X-ray $E F_E$ flux at 100 keV measured in the soft states of Cyg X-3 is $\approx$0.2 keV cm$^{-2}$ s$^{-1}$ \citep{SZM08}. The observed $\sim$100\,keV flux exhibits strong orbital modulation \citep{Zdziarski12b} similar to that at lower energies, with the minimum around the superior conjunction, which could be due to wind absorption. This implies that most of the X-ray emission originates in the accretion flow rather than in the jet, whose emission shows a very different modulation pattern. On the other hand, the duration of the flaring state is by a factor of $\sim$7 shorter than the one of the soft state. Thus, we require that jet contribution at 100 keV to be $<$0.1 keV cm$^{-2}$ s$^{-1} \approx 1.6 \times 10^{-10}$ erg cm$^{-2}$ s$^{-1}$.

Furthermore, the model synchrotron flux in the flaring state cannot exceed available measurements. We use the observational constraint from the Submillimeter Array (SMA) observations during flaring periods in 2018, with the average flux of $\sim$300\,mJy at 225\,GHz \citep{McCollough23}. That emission undergoes free-free absorption on electrons of the stellar wind. This process can influence both the observed level of 225\,GHz emission and the profile of its orbital modulation. We estimate the distance away from the star at which the optical depth of the free-free absorption of 225\,GHz emission drops to unity. Using typical values for the mass loss rate of the donor, velocity and temperature of the stellar wind in Cyg X-3 (e.g., \ant), and assuming a fully ionized helium plasma, we find this critical distance to be $\sim 20 a$. This provides an approximate constraint on the location of the region responsible for producing the 225\,GHz emission.

\subsection{The models}
\label{models}

We consider two models.

Model 1. The jet is assumed to have a fixed orientation. The dominant radiative process is the anisotropic Compton scattering of stellar blackbody photons in a compact region along the jet. The orbital modulation is entirely due to this process since the Doppler boosting does not depend on the orbital phase. As found before (\citealt{Dubus10}; \aaz), this model fits well the \g-ray folded light curve, while {\it requiring\/} the jet to be significantly misaligned with respect to the orbital axis. Since the jet also contains magnetic field, the SSC contributes to the folded light curve at a constant level, which effect was not calculated before (except for a qualitative consideration in \citealt{Zdziarski12a}). Here, we calculate the maximum allowed field strength allowed by the data. The main parameters of this model are the jet angles and the height of the emission region along the jet. The emission of both the jet and the counterjet are taken into account, and we have
\begin{equation}
\varphi_{\rm j} = {\rm const}, \quad {\mathbfit e}_{\rm cj} = - {\mathbfit e}_{\rm j}, \quad \varphi_{\rm cj} = \pi+\varphi_{\rm j}, \quad \theta_{\rm cj} = \pi-\theta_{\rm j}.
\label{offset1}
\end{equation}

Model 2. The jet average direction is aligned with the orbital axis but the thrust of the stellar wind bends the jet outward by an angle $\theta_{\rm j}$ along the line connecting the jet with the center of the donor \citep{Yoon15, Yoon16, Bosch-Ramon16, Barkov22}. This effect alone would result in a jet precessing at the orbital period with the azimuth given by $\varphi_{\rm j} = \varphi$. However, the Coriolis force arising from the orbital motion will induce a lateral bending of the jet opposite to the direction of the rotation, leading to formation of a helix. Here, we are considering emission of a compact region of the jet, which can be then offset by an arbitrary angle, $\varphi_{\rm o}$. We note that the Coriolis force acts in the same direction on the jet and counterjet, which is evident upon considering the relevant vector product. This implies that both jet and counterjet experience lateral bending opposite to the direction of rotation. Then, we have 
\begin{align}
& \varphi_{\rm j} = \varphi-\varphi_{\rm o}, \\
& {\mathbfit e}_{\rm cj} = (\sin \theta_{\rm j} \cos \varphi_{\rm j},  \sin \theta_{\rm j} \sin \varphi_{\rm j}, -\cos \theta_{\rm j}), \\
& \varphi_{\rm cj} = \varphi_{\rm j}, \quad \theta_{\rm cj} = \pi-\theta_{\rm j}.
\label{offset2}
\end{align}
The variable $\varphi_{\rm j,cj}$ results then in a variability of the Doppler boosting, which is then responsible for the orbital modulation. We note that our treatment of the geometry is approximate, since we assume a straight jet connecting the origin with the emission region, while the jet actually has a helical shape. The exact details of this helical shape are influenced by the jet acceleration profile, i.e., how the jet acceleration rate (and hence the jet velocity) evolves along the length of the jet. Since we do not have precise information on this acceleration profile, we have to neglect this complication.

A recent study by \cite{Yang23} suggests the presence of a non-negligible transversal component in the jet of Cyg X-3, albeit observed in the hard state. Nevertheless, it is possible that this transversal component may arise from the wind-induced jet precession, and hence the apparently broadened jet in those observations may be attributed to its helical nature, supporting our assumptions about the jet morphology in the Model 2.

The jet will still be irradiated by the stellar blackbody, which adds a perturbation to the modulation due to the variable Doppler boosting of the precessing jet. We calculate the maximum flux, $\propto R_*^2 T_*^4$, of the blackbody allowed by the model for a given $h$, constraining in turn the radius and temperature of the donor. The maximum depends on the height of the emission region, $h$.

We assess the significance of the attenuation of \g-rays en route to the observer as they traverse the photon field of the donor star and suffer \g--\g absorption. Our analysis shows that this effect is entirely negligible for the jet, with substantial opacity emerging only beyond $\sim$100 GeV. For the counterjet, the onset of significant attenuation is, however, found to arise already at 5--6 GeV for Model 1, and 10--15 GeV for Model 2. Still, considering its subdominant contribution to the total emission and the increasingly large uncertainties in the data beyond these energies, we conclude that, in our specific case, this effect can be neglected also for the counterjet.

With either model, we conduct fitting of the phase-averaged spectrum and the light curve. We first obtain the spectrum of the emission as a function of the orbital phase for assumed preliminary values of $\theta_{\rm j}$, either $\varphi_{\rm j}$ or $\varphi_{\rm o}$, $\beta_{\rm j}$, $i$, and $h$. With that, we fit the observed phase-averaged photon spectrum, obtaining $E_{\rm e}$, $p$, $\gamma_{\rm min}$ and $\gamma_{\rm max}$. With those, we fit in turn the observed folded light curve, which yields updated values of the parameters. Using those more accurate estimates, we fit again the phase-averaged spectrum, iterating in this way until convergence. The fitting is performed using the python \texttt{LMFIT} module (employing the default Levenberg-Marquardt minimization method). 

\section{Results for Model 1: Blackbody Compton}
\label{bbc}

\subsection{Fit results}
\label{bbc_fit}

\setlength{\tabcolsep}{3pt}
\begin{table*}
\caption{Fit results for the phase-averaged spectrum and the light curve for Model 1 without and with magnetic field \label{tab:paramsstablej}
}
\vskip -0.4cm
\centering
\begin{tabular}{cccccccccccccccc}
\hline
$\alpha_{\rm j}(\degr)$ & $B_{\rm max}$(G) &$h/a$ & $i(\degr)$ & $\theta_{\rm j}(\degr)$ & $\varphi_{\rm j}(\degr)$ &  $\beta_{\rm j}$ & $\chi_\nu^2$ & $p$ & $\gamma_{\rm min}\,(10^3)$ & $E_{\rm e}\,(10^{37}$erg) & $\chi_\nu^2$ & $\gamma_{\rm br}$ & $i_{\rm j}(\degr)$ & $\Delta\lambda(\degr)$ \\
\hline
-- & 0 & $1.96 \pm 0.09$ & $28 \pm 6$ &$32 \pm 7$ &$188 \pm 2$ &$0.59 \pm 0.05$ & 7.2/5 & $4.16 \pm 0.05$ & $2.9 \pm 0.2$ & $18 \pm 1$ & 5.1/7 & 426  & $5_{-3}^{+12}$ & $228_{-35}^{+118}$\\
5 & 34 & $2.3 \pm 0.6$ & $33 \pm 7$ & $35 \pm 8$ &$188 \pm 3$ &$0.7 \pm 0.2$ & 7.3/5 & $4.8 \pm 0.1$ & $3.4 \pm 0.4$ & $14 \pm 2$ & 7.5/7 & 591 & $5_{-2}^{+12}$ & $261_{-67}^{+88}$\\
\hline
\end{tabular}\\
\tablecomments{The Lorentz factor of the break, $\gamma_{\rm br}$, jet viewing angle, $i_{\rm j}$, and the difference in the position angles, $\Delta\lambda$, are derived quantities rather than free parameters. In all cases, $\gamma_{\rm max}=10^6$. The value of $B_{\rm max} = 34$ G corresponds to the upper limit for the jet opening angle of $\alpha_{\rm j} = 5\degr$, while the fit does not depend on $\alpha_{\rm j}$ for $B=0$.
}
\end{table*}

\begin{figure*}
     \centering
\begin{overpic}[height=0.32\textwidth]{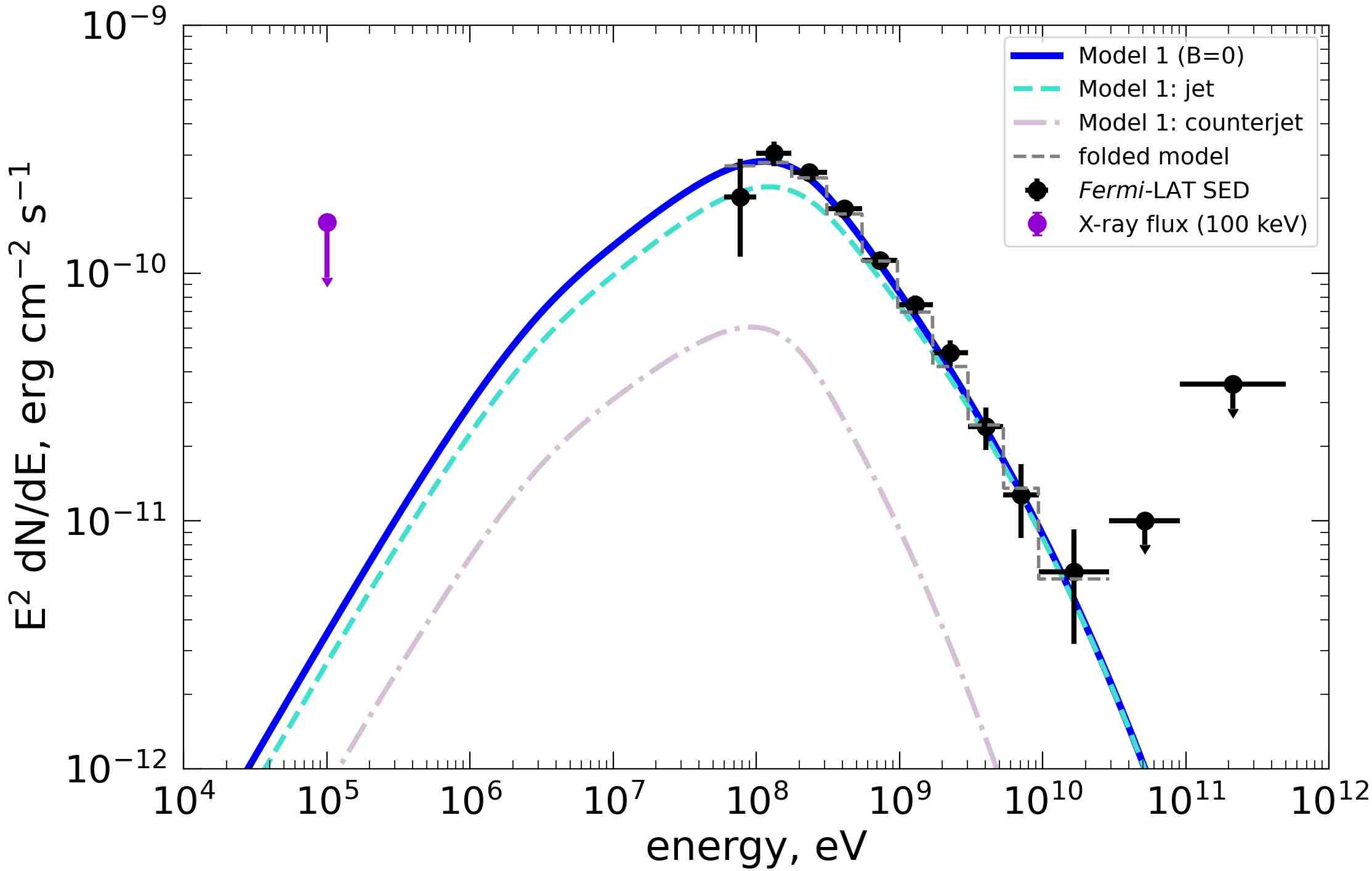}
 \put(0.08\textwidth, 0.285\textwidth){\large (a)}
 \end{overpic}
 \hfill
\begin{overpic}[height=0.32\textwidth]{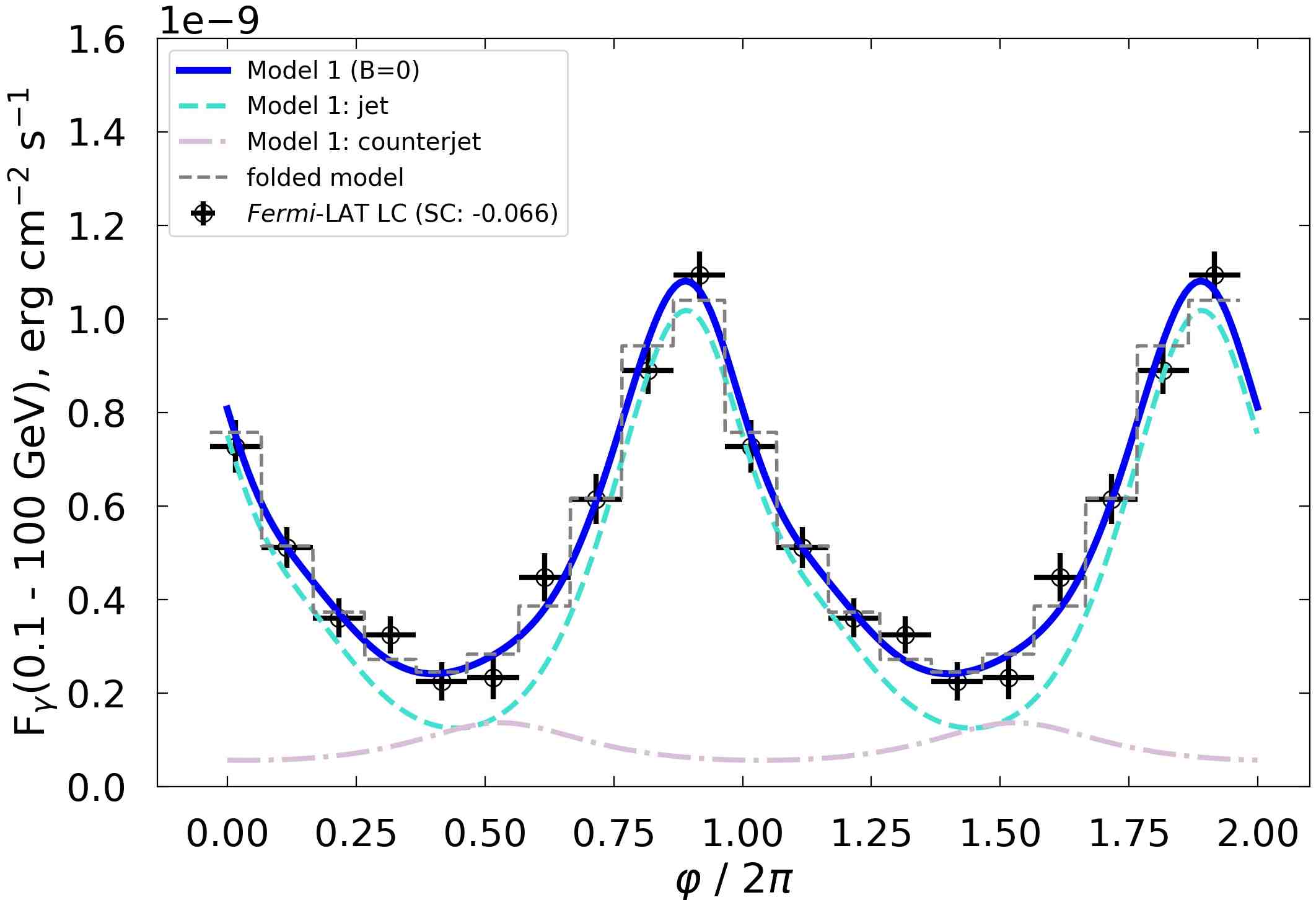} 
 \put(0.41\textwidth, 0.28\textwidth){\large (b)}
 \end{overpic} \\
\begin{overpic}[height=0.32\textwidth]{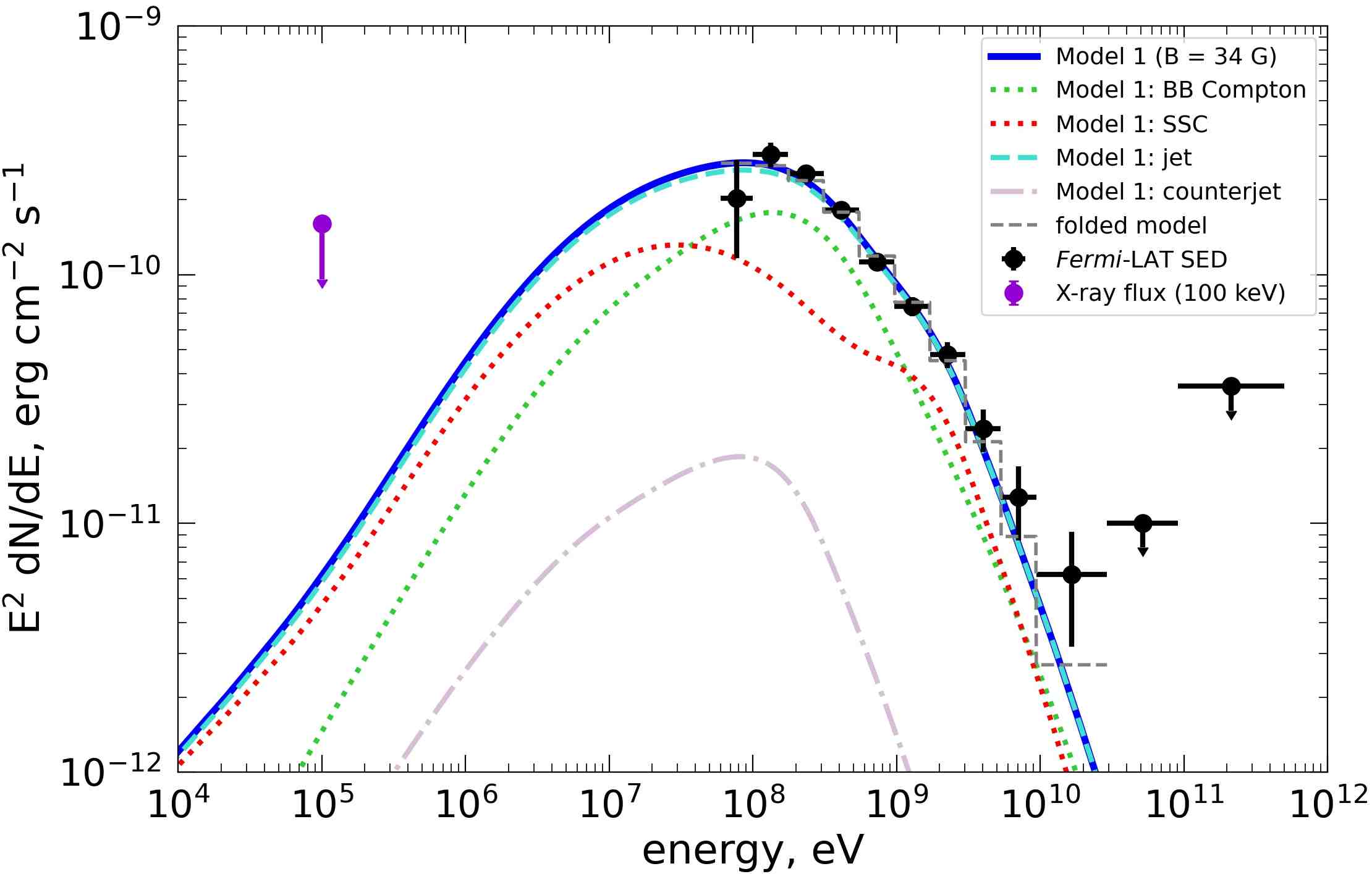}
 \put(0.08\textwidth, 0.285\textwidth){\large (c)}
 \end{overpic}
 \hfill
\begin{overpic}[height=0.32\textwidth]{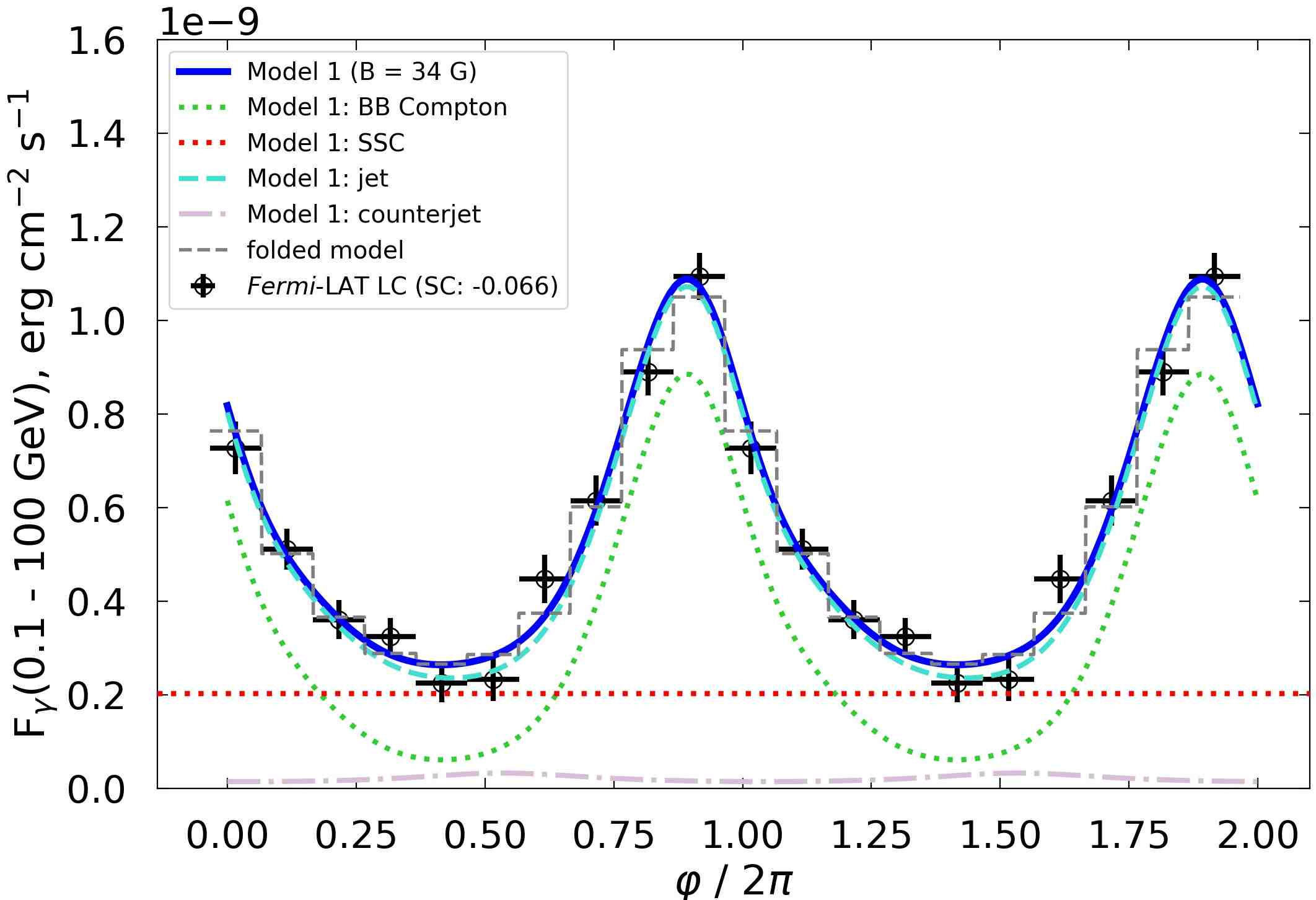}
 \put(0.41\textwidth, 0.28\textwidth){\large (d)}
 \end{overpic}
        \caption{Fit results for Model 1 without and with magnetic field. Black error bars show the LAT data, the purple SED data point (upper limit) shows the constraint on the X-ray flux at 100 keV of $\approx 1.6 \times 10^{-10}$ erg cm$^{-2}$ s$^{-1}$, the dark blue solid curves show the best-fit models, and the dashed histograms show the model predictions for the binned fluxes. Cyan dashed and the grey dash-dotted lines display the contributions from the jet and counterjet, respectively. (a) The phase-averaged spectrum for $B=0$. (b) The 0.1--100\,GeV light curve folded and averaged over the orbital phase with respect to the superior conjunction, $\varphi$. (c) Similar to (a), but for the upper limit value of the magnetic field $B_{\rm max} = 34$ G. (d) The corresponding light curve. In (c) and (d), the light-green and red dotted curves show the contributions from the Compton scattering of blackbody photons and the SSC, respectively.  }
        \label{fig:fitstablejet}
\end{figure*}

We assume $R_*=10^{11}$\,cm and $T_*=10^5$\,K \citep{Koljonen17}. Notably, that study also reveals that the photosphere extends significantly beyond the WR star, implying that the \g-ray emitter might reside within an optically thick wind. This would then influence the blackbody Compton emission, as well as the \g--\g\ absorption, adding complexity to our modeling approach. To address this, we make a simplifying assumption that the jet partially clears its surrounding environment, pushing the optically thick wind to further distances. Thus, we neglect this complication, which was also done in the previous studies (\citealt{Dubus10}, \aaz). We first include only Compton scattering of blackbody photons. Our results are shown in Figures \ref{fig:fitstablejet}(a,b) and Table \ref{tab:paramsstablej}. Confirming the previous results, e.g., \aaz, we find the steady state electron distribution to be steep, with $p \approx 4.2 \pm 0.1$, with a large low-energy cutoff, $\gamma_{\rm min} \approx (2.9 \pm 0.2)\times 10^3$, and the emission height of $h\approx (2\pm 0.1)a$. We find that the high-energy cutoff is not required, and set it to a high value of $\gamma_{\rm max} = 10^6$. This implies that the decline due to KN effects alone is sufficient to explain the steepening at high energies in the observed spectrum. We show the folded light curve fitted by our model vs.\ $\varphi$, i.e., the phase with respect to the actual superior conjunction, assuming the best fit offset of \ant. We see that the peak of the modulation is still significantly before the superior conjunction, implying that the jet is inclined, see Table \ref{tab:paramsstablej}. The value of the $\varphi_{\rm j}>180\degr$ means that the jet is slightly bent in the direction of the rotation around the superior conjunction, which causes the peak of the scattered flux to occur before it. The jet viewing angle is low, $i_{\rm j}\approx 5\degr^{+12}_{-3}$, which is in agreement with radio constraints \citep{Mioduszewski01, Miller-Jones04}. The projection of the orbital axis on the sky shows a large offset with respect to that of the jet, $\Delta\lambda \approx 228\degr^{+118}_{-35}$. 

An earlier study by \citet{Zdziarski12a} found the magnetic field strength limited to $B \lesssim 100$\,G in the \g-ray emitting region of the jet, obtained relying only on the spectral measurements. Here, we impose stringent constraints on the magnetic field strength employing the current \g-ray data with the spectrum and the light curve, an upper limit for the X-ray flux at 100 keV, and the millimeter measurements. 

We have four independent constraints from the data that allow us to infer the maximum allowed magnetic field strength. The first follows from the \g-ray modulation light curve, for which the SSC flux cannot exceed its minimum value. Then, the SSC contribution cannot overproduce the emission at $\lesssim 0.1$ GeV. Depending on the magnetic field, this low-energy emission can become quite strong given the presence of cooled low-energy electrons below $\gamma_{\rm min}$ (see Section \ref{electrons}). The remaining two constraints are described in Section \ref{observational}, and are given by the observed 100\,keV and 225\,GHz fluxes, which cannot be exceeded by the model spectrum. 

Following this approach, we derive an upper limit for $B$ using the jet opening angle $\alpha_{\rm j}=5\degr$, a value found on the scale of $\sim\! 10^{15}$\,cm ($5.0\pm 0.5\degr$; \citealt{Miller-Jones06}). Specifically, we obtain $B_{\rm max} \approx 34$ G. This magnetic field strength yields an SSC contribution saturating the minimum of the light curve, while we find that the X-ray and synchrotron emissions remain significantly below the observational constraints. Figures~\ref{fig:fitstablejet}(c,d) illustrate the fits to the \g-ray data. The resulting fit parameters are given in Table \ref{tab:paramsstablej}, and are similar to those obtained assuming zero magnetic field.

The minimum length of the emitting region can be set by $h_{\rm cool}\sim t_{\rm cool}(\gamma_{\rm min}) \beta_{\rm j} c$, where the cooling time, $t_{\rm cool}(\gamma)\equiv \gamma/|\dot{\gamma}_{\rm cool}|$, depends on the total cooling rate, $\dot{\gamma}_{\rm cool}$. We calculate $\dot{\gamma}_{\rm cool}$ due to the stellar Compton process, $\dot{\gamma}_{\rm cool, SC}$, using eq.~16 in \citet{Zdziarski14a}, while the cooling rate due to the SSC (for which the KN effects are negligible) and the synchrotron processes are evaluated as $|\dot{\gamma}_{\rm cool, SSC/syn}| = (4 \sigma_{\rm T} \gamma^2 U_{\mathrm{syn}/B})/(3 m_{\rm e} c)$, where $U_{\mathrm{syn}/B}$ is the energy density of the target synchrotron emission and of the magnetic field, respectively, and $m_{\rm e}$ is the electron mass. For the model with the maximum magnetic field we obtain $t_{\rm cool}(\gamma_{\rm min}) \approx 7.7$ s. This implies $h_{\rm cool} \approx 1.6 \times 10^{11}\,{\rm cm}\approx 0.25 h < \Delta h= h/3$, as required (since the dissipation region is likely to be extended). Next, we evaluate the angle-integrated jet bolometric luminosity for the same case, obtaining $L_{\rm j} \approx 2.4 \times 10^{37}$ erg s$^{-1}$. The ratio $E_{\rm e}/L_{\rm j}$ gives the average energy loss time scale of all electrons. We find that this time scale is similar to the cooling one at $\gamma_{\rm min}$. 

\subsection{Caveats for  Model 1}
\label{against}

\begin{figure}
\includegraphics[width=\columnwidth]{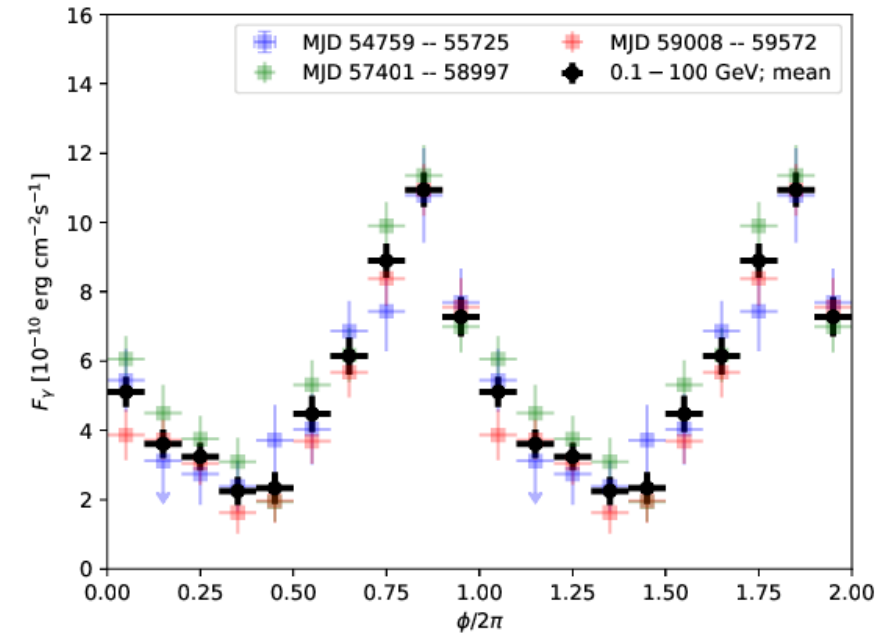}
\centering
\caption{Comparison of the orbital modulation profiles during different time intervals. We see the profiles remain approximately constant during $\sim$5000 days of the LAT monitoring. Here, the horizontal axis gives the phase, $\phi$, according to the ephemeris of \citet{Antokhin19}.
}
\label{intervals}
\end{figure}

\begin{figure}
\includegraphics[width=\columnwidth]{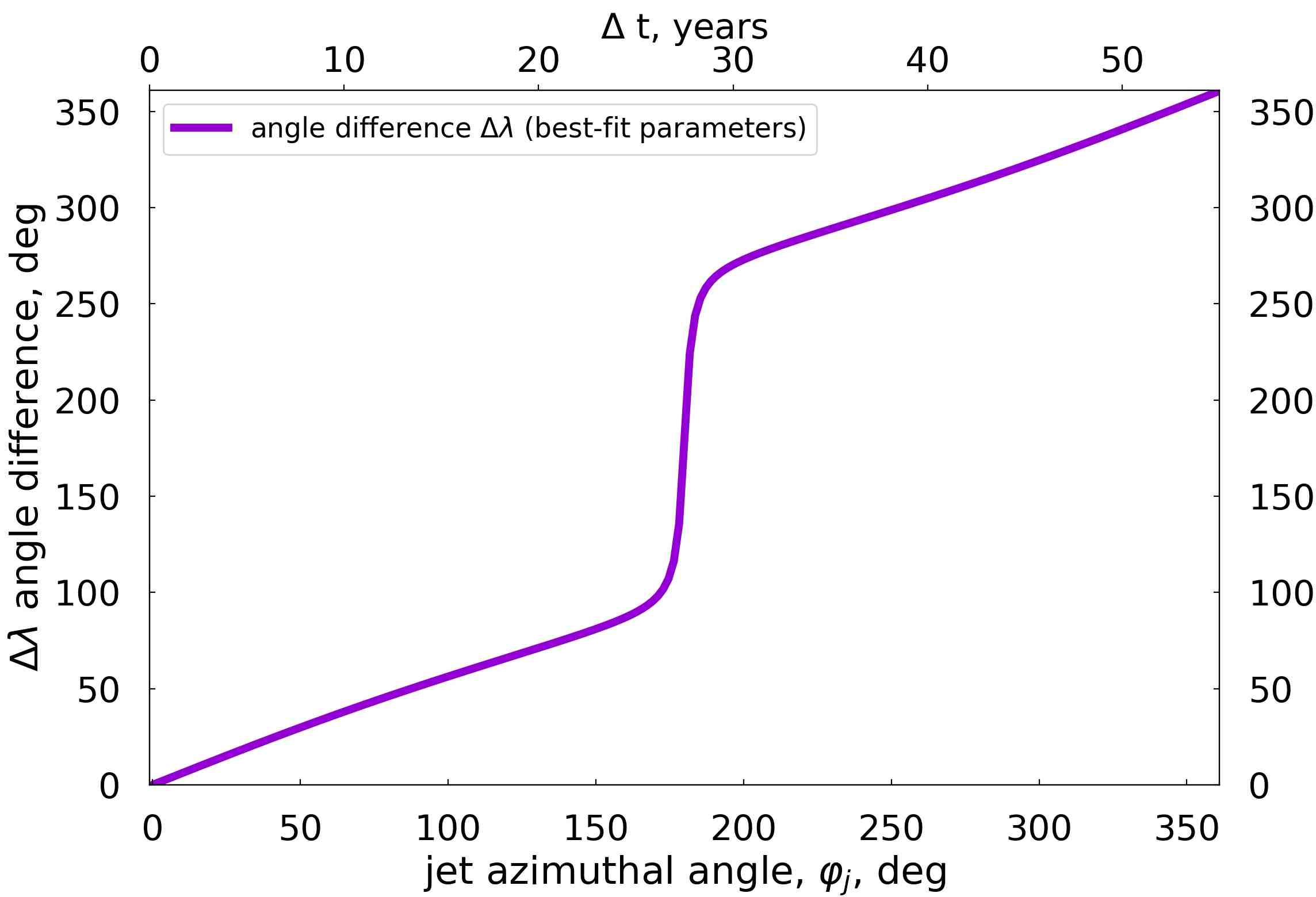}
\centering
\caption{The predicted angular difference between the projections on the sky of the jet and the orbital axis, $\Delta \lambda$, see Equations (\ref{delta}) and (\ref{precession}), for the case of a jet aligned with the spin axis of the compact object but misaligned with respect to the orbital axis. The jet azimuthal angle, $\varphi_{\rm j}$, is predicted to change by $360\degr$ during the precession period of 50\,yr, which is clearly not seen. We assumed the best-fit values of $\theta_{\rm j}$, $\phi_{\rm j}$ and $i$ for the case with the maximum magnetic field$B_{\rm max} = 34$ G. Currently, that model gives $\varphi_{\rm j}\approx 188\degr$, $\Delta\lambda \approx 261\degr$, see Table \ref{tab:paramsstablej}.}
\label{fig:angl_diff}
\end{figure}

If the jet axis is linked to the rotation axis of the compact object, prograde geodetic precession is predicted. The De Sitter precession period for either a BH or a neutron star is given by \citep{Barker75, Apostolatos94},
\begin{equation}
P_{\rm prec}=\frac{c^2(M_*+M_{\rm c})^{4/3} P^{5/3}}{(2\pi G)^{2/3}(2+3 M_*/2 M_{\rm c})M_* M_{\rm c}},
\label{precession}
\end{equation}
where the Kepler law was used, $M_{\rm c}$ includes the contribution from the rotational energy, and $G$ is the gravitational constant. For the best-fit masses of \ant, $P_{\rm prec}\approx 50\,{\rm yr}\approx 18 000$\,d, relatively weakly depending on them. During $P_{\rm prec}$, the jet azimuth, $\varphi_{\rm j}$, changes by $360\degr$. The observations of \g-rays by the LAT covered already $\approx 5 000$\,d (see Figure \ref{lc}), so we would expect substantial changes in the modulation profile. However, we see no evidence for such changes, as shown in Figure \ref{intervals}. 

Another argument against the presence of the precession is from the beat between the orbital and precession periods. The two angular frequencies add, leading to a shift of the frequency of the observed orbital modulation of the jet \g-ray emission. This shift is $\approx$0.2\,s, while the \g-ray period measured by us equals the orbital one with the accuracy of $<0.02$\,s, see Section \ref{Data}.

A consequence of the geodetic precession is also a full rotation of the jet position angle with $P_{\rm prec}$. The predicted values of $\Delta\lambda$ are shown in Figure \ref{fig:angl_diff}. However, we see no evidence for such changes in radio observations from 1985 to 2016 \citep{Molnar88, Schalinski95, Mioduszewski01, Miller-Jones04, Egron17}. The position angle of the approaching jet remains relatively constant at close to $180\degr$ with respect to the north. 

Furthermore, this model predicts no Doppler orbital modulation of the synchrotron emission, since it is coming from a jet with a fixed orientation. This may disagree with observations if the preliminary results showing strong orbital modulation at 225\,GHz similar to those of the \g-rays \citep{McCollough23} are confirmed.

The jet could possibly be misaligned due to some asymmetry in the binary, e.g., the presence of a bow shock (\ant), instead of being in the direction of the (misaligned) BH spin. However, strong X-ray linear polarization has been discovered from this system \citep{Veledina24, Veledina24b}. The high polarization degree and the polarization angle perpendicular to the position angle of the radio jet demonstrate we do not see the primary X-ray source directly but instead we see only the X-rays scattered in the inner an accretion funnel (expected for super-critical accretion). This also shows that the funnel, located close to the BH, is aligned with the jet. Thus, a constraint on this origin of the misalignment is that it has to already start on that scale.  

\section{Results for Model 2: orbital precession}
\label{precessing_jet}

\subsection{Fit results}
\label{prec_fit}

\setlength{\tabcolsep}{3.4pt}
\begin{table*}
\caption{Fit results for the phase-averaged spectrum and the light curve for a bent and precessing jet (Model 2, pure SSC) for the maximum value of the magnetic field, $B = 100$ G. \label{tab:fits_precj_ssc_maxb}
}
\vskip -0.5cm
\centering
\begin{tabular}{cccccccccccccccc}
\hline
$B$(G) & $\alpha_{\rm j}(\degr)$ & $h/a$ & $i(\degr)$ & $\theta_{\rm j}(\degr)$ & $\varphi_{\rm o}(\degr)$ &  $\beta_{\rm j}$ & $\chi_\nu^2$ & $p$ & $\gamma_{\rm min}$ ($\times 10^3$) & $\gamma_{\rm max}$ & $E_{\rm e}\,(10^{38}$erg) & $\chi_\nu^2$ & $\gamma_{\rm br}$ & $F_{\rm X}$ & $\beta_{\rm eq}$ \\
\hline
100 & 5(f) & 11(f) & 30(f) & $41 \pm 2$ & $142 \pm 4$ & $0.46 \pm 0.02$ & 10.3/7 & $3.97 \pm 0.06$ & $1.1 \pm 0.2$ & $10^6$(f) & $24 \pm 3$ & 8.1/7 & 95 & 2.8 & 30.7 \\
\hline
\end{tabular}\\
\tablecomments{$F_{\rm X}$ is the model-predicted $E F_E$ flux at 100 keV in units of $10^{-10}$ erg cm$^{-2}$ s$^{-1}$, and (f) denotes a fixed parameter. 
}
\end{table*}

\begin{figure*}
     \centering
\hspace*{-5mm} \begin{overpic}[height=0.31\textwidth]{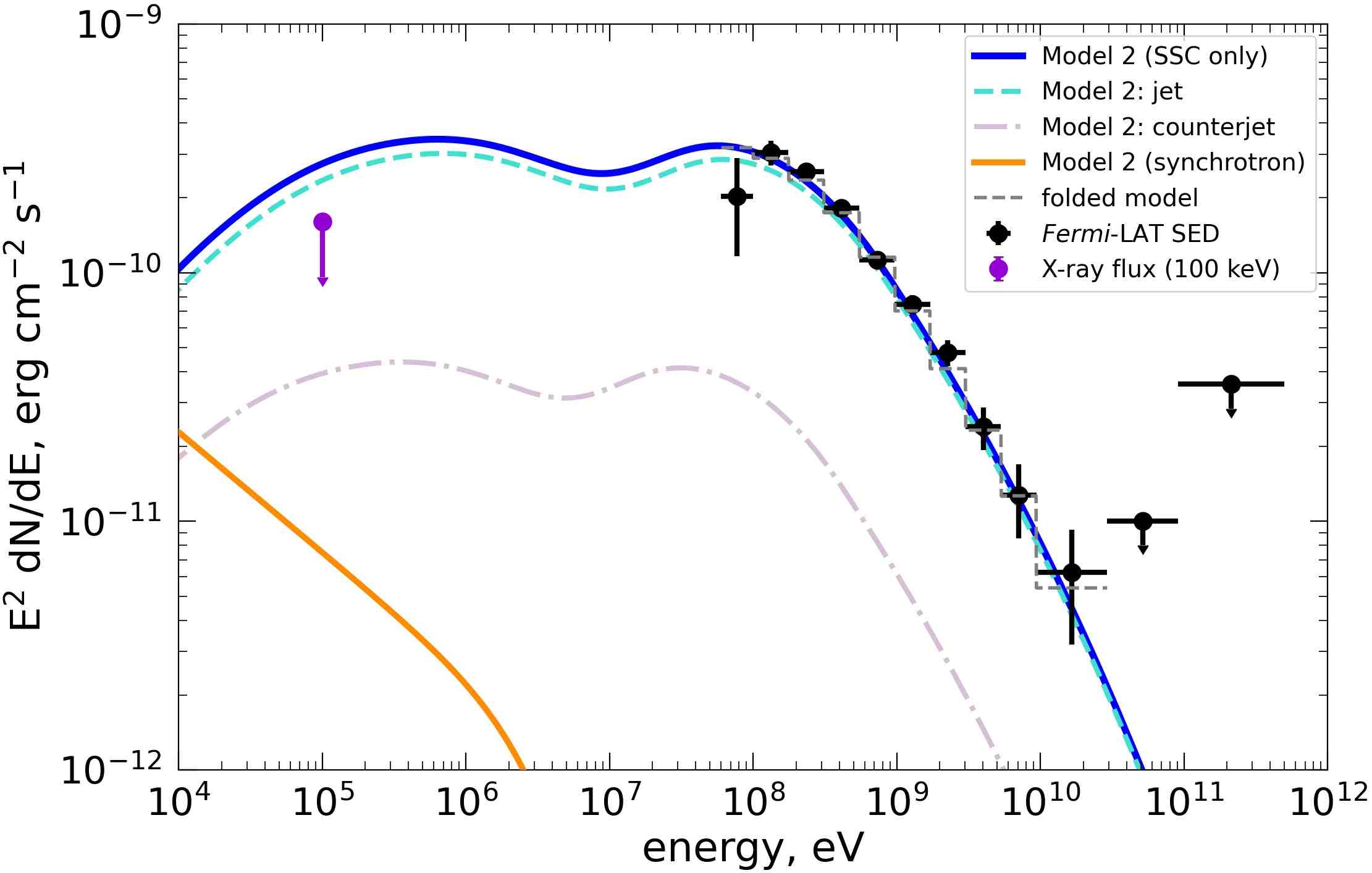}
 \put(0.08\textwidth, 0.275\textwidth){\large (a)}
 \end{overpic}
 \hspace*{6mm}
\begin{overpic}[height=0.31\textwidth]{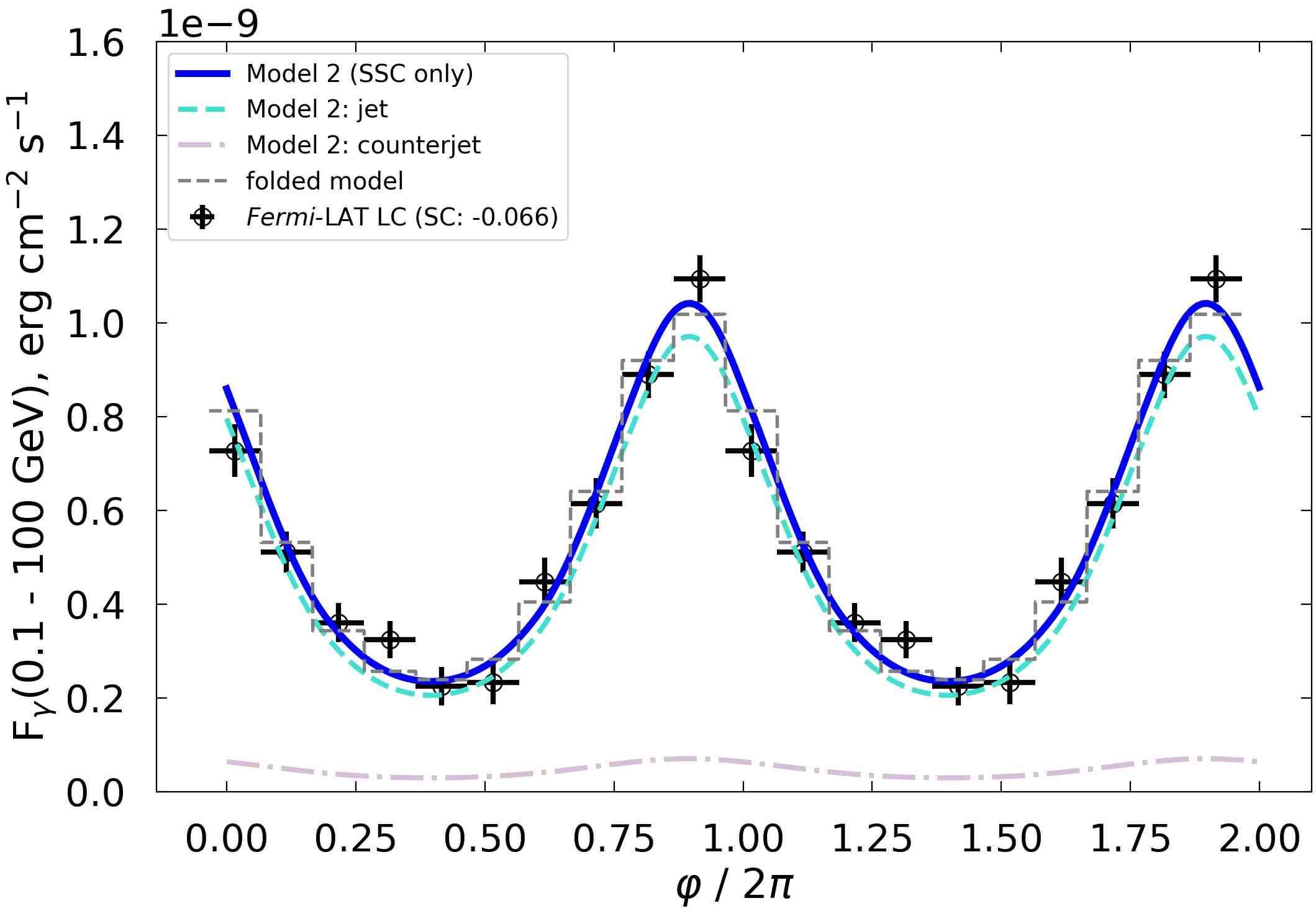} 
 \put(0.2\textwidth, 0.27\textwidth){\large (b)}
 \end{overpic} \\
\begin{overpic}[height=0.3\textwidth]{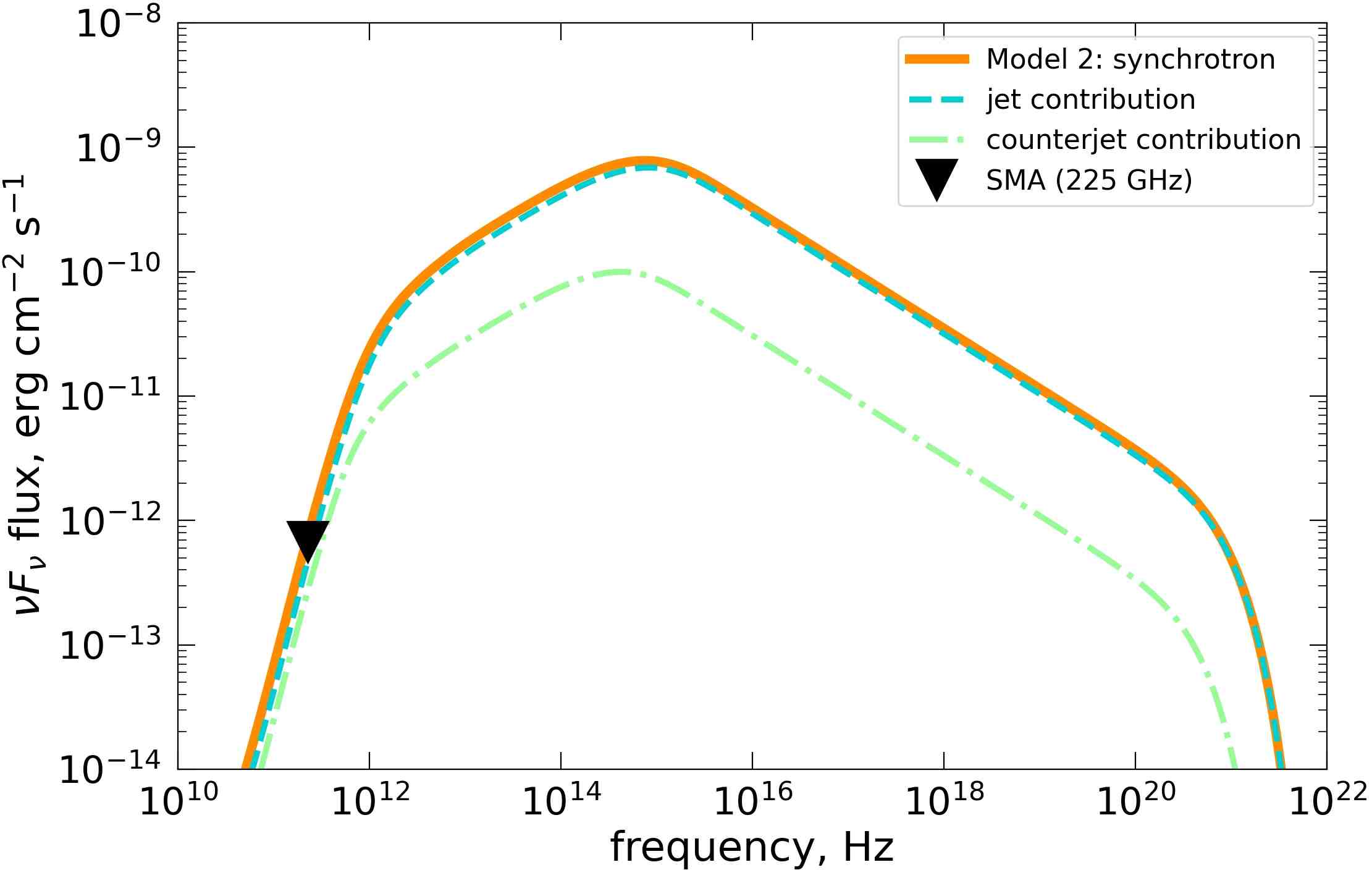}
 \put(0.08\textwidth, 0.264\textwidth){\large (c)}
 \end{overpic}
 \hspace*{3mm}
\begin{overpic}[height=0.3\textwidth]{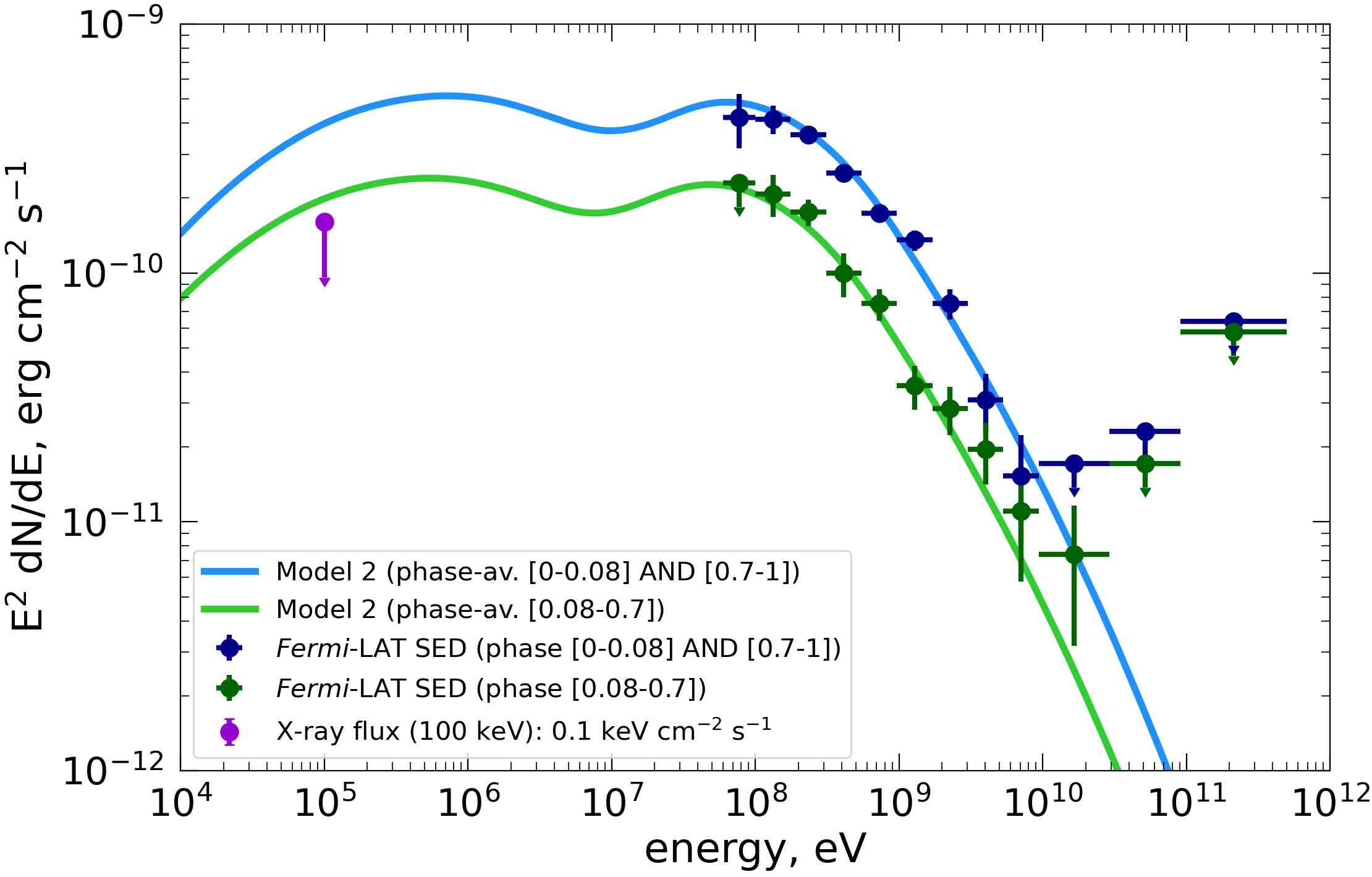}
 \put(0.08\textwidth, 0.264\textwidth){\large (d)}
 \end{overpic}
        \caption{Best-fit solutions for the case of the jet precessing at the orbital period, for the derived upper limit for the magnetic field, $B_{\rm max} = 100$ G. (a) The phase-averaged spectrum and (b) the 0.1--100\,GeV light curve.  The blue solid curves represent the best-fit model, the dashed histograms show the model predictions for the binned fluxes, and the jet and counterjet contributions are depicted with the dashed cyan and dash-dotted grey curves, respectively. (c) The predicted contribution from the synchrotron emission along with the average flux from the SMA, shown by the black triangle. The maximum magnetic field corresponds to the model 225 GHz flux equal to the observed one. The jet and counterjet contributions are shown with dashed turquoise and dash-dotted green curves, respectively. (d) Comparison of the {\it Fermi}-LAT data and the phase-averaged models for the data split into two phase parts corresponding to the flux higher and lower than $6 \times 10^{-10}$ erg cm$^{-2}$ s$^{-1}$ level (shown in blue and green, respectively). }
        \label{fig:fitprecjetmaxb}
\end{figure*}

In this model, the synchrotron and SSC emissions are orbitally modulated due to the varying Doppler factor associated with the variable viewing angle of the jet. The emission peaks at the lowest jet viewing angle, which occurs at $\varphi = \pi + \varphi_{\rm o}$ (at the inferior conjunction for $\varphi_{\rm o} = 0$). Since the Doppler boosting is independent of the emission height, the fitted SSC spectrum and modulation profile do not depend on $h$. However, the synchrotron flux does depend on it, see Equation (\ref{ratio}). Since both the SSC and synchrotron are similarly modulated due to this process, this model could explain the preliminary report of a 225\,GHz orbital modulation \citep{McCollough23}.

When permitting all parameters in the light curve fit to vary independently, we observe significant uncertainties in their inferred values, indicating the presence of degeneracies among various parameters. Specifically, the same jet viewing angle $i_{\rm j}$ and hence the same Doppler factor modulation profile can be achieved with different combinations of $\theta_{\rm j}$ and $i$, implying a correlation between these parameters. To address this, we opt to fix $i = 30\degr$ (\ant).

As previously, we use $\alpha_{\rm j}=5\degr$ \citep{Miller-Jones06}. We search for solutions that also reproduce the average flux measured at 225 GHz \citep{McCollough23}. We find that configurations assuming a predicted position of the jet--wind recollimation shock of $h\sim a$ (e.g., \citealt{Yoon16}) significantly underpredict the 225\,GHz flux, regardless of the magnetic field strength. Consequently, we consider $h/a \gg 1$, which increases the synchrotron flux, see Equation (\ref{ratio}). We  find $h=11 a$ and $B=100$ G as the (approximate) minimum distance of the emitting region and the maximum magnetic field strength, respectively, yielding the observed flux at 225\,GHz. The fit parameters for this $B$ are given in Table \ref{tab:fits_precj_ssc_maxb} and the spectrum and the light curve are shown in Figures \ref{fig:fitprecjetmaxb}(a,b). The corresponding model prediction for the phase-averaged synchrotron emission is shown in Figure \ref{fig:fitprecjetmaxb}(c).

We then compare our solution to the dependence of the total magnetic field strength in the comoving frame on the distance, $B(h)$, of the analytical jet model of \citet{ZSSN22}. For it, we assume the jet power of $P_{\rm j}= 9\times 10^{38}$\,erg\,s$^{-1}$, the terminal magnetization parameter of $\sigma_{\rm min} = 0.008$, where $\sigma \equiv (B^2/4\pi)/(\rho c^2)$, $\rho$ is the comoving mass density, and the spin parameter of $a_*=1$ and other parameters defined in that paper of $\ell=0.5$, $a_r=1$, $a_\Gamma=0.2$, $q_2=1.9$ where assumed. They yield the jet half-opening angle of $\alpha_{\rm j}\approx 5\degr$ and $B\approx 100$\,G (dominated by the toroidal component) at $h/a = 11$. 

In addition, we generate model predictions for the average spectra corresponding to the {\it Fermi}-LAT high and low-flux states, delineated by the 0.1--100 GeV light curve flux being above and below the threshold of $6 \times 10^{-10}$ erg cm$^{-2}$ s$^{-1}$, respectively. This threshold approximately corresponds to the average between the maximum and minimum fluxes observed in the light curve. This partitioning yields two distinct phase intervals: (1) $0 \leq \varphi/(2\pi) < 0.08 \ \cup 0.7 \leq \varphi/(2\pi) < 1$ where the 0.1 - 100 GeV flux is above the separation threshold, and (2) $0.08 \leq \varphi/(2\pi) < 0.7$ where the flux is below the threshold. The comparison of the two spectra with the model predictions is shown in Figure \ref{fig:fitprecjetmaxb}(d). We see no noticeable spectral changes between the two substates, which are also well described by our model. 

Furthermore, we estimate upper limits on $R_*$ and $T_*$ by including the blackbody Compton component into the model. Based on observations and stellar modeling, \citet{Koljonen17} found $T_* \approx 0.8$--$1.0\times 10^5$\,K at the range of $R_*\approx 1.5$--$1.0\times 10^{11}$\,cm, corresponding to $R_* T_*^2\approx 1.0\times 10^{21}$\,cm\,K$^2$. Interestingly, we find a non-monotonic behavior in the $\chi^2$ of the light curve fit as a function of $R_* T_*^2$; it is initially decreasing, then increasing again. The best fit is achieved for a value of $R_* T_*^2 = 1.6 \times 10^{21}$\,cm\,K$^2$ with $\chi_{\nu}^2 = 7.8/7$ ($\Delta \chi^2 = 2.5$ with respect to the pure SSC case). However, the $\chi^2$ of the spectral fit increases monotonically with increasing $R_* T_*^2$, with an increase by $\Delta \chi^2 = 2.7$ as compared to the pure SSC fit being achieved coincidentally at the same value of $R_* T_*^2$ ($\chi_{\nu}^2 = 10.8/7$). Given the best fit of the light curve and a still satisfactory SED fit, this value of $R_* T_*^2$ can be therefore considered as the most optimal one. The best-fit range for $R_* T_*^2$ for this most optimal solution found by imposing $\Delta \chi^2 \leq 2.7$ for the spectral fit (as it degrades much quicker), is $R_* T_*^2= (0$--$1.7) \times 10^{21}$\,cm\,K$^2$. The higher $h$, the higher the allowed range of $R_* T_*^2$. In addition, within this best-fit range, the model-predicted X-ray flux at 100 keV shows a slow decrease with an increasing $R_* T_*^2$, thus having consistently lower values compared to the pure SSC case (due to adjusted $E_{\rm e}$ in the fit). We find the X-ray flux at 100 keV for $R_* T_*^2 = 1.6 \times 10^{21}$\,cm\,K$^2$ of $\approx 1.4 \times 10^{-10}$ erg cm$^{-2}$ s$^{-1}$, which is below the observational constraint.

\subsection{Jet parameters}
\label{parameters}

The model of \citet{ZSSN22} gives the magnetic flux threading the black hole as a function of $P_{\rm j}$ in their equation (17), which is based on numerical results of \citet{Tchekhovskoy09}. It yields $\Phi_B\approx 5\times 10^{21}$\,G\,cm$^2$. On the other hand, since Cyg X-3 appears to be a super-Eddington accretor \citep{Veledina24}, $\dot M_{\rm accr}\gtrsim 2\times 10^{19}$\,g\,s$^{-1}$ (using the Eddington luminosity for He and the accretion efficiency of 0.1). The magnetic flux at the limit of the magnetically arrested disk (MAD; \citealt{BK74,Narayan03}) is $\Phi_{\rm MAD}\approx 50 (\dot M_{\rm accr} c)^{1/2}r_{\rm g}$ \citep{Tchekhovskoy11}, implying $\Phi_{\rm MAD}\gtrsim 4\times 10^{22}$\,G\,cm$^2$. Thus, the jet has the power at least an order of magnitude below the MAD limit.

We then find the electron cooling time at $\gamma_{\rm min}$ of $t_{\rm cool}(\gamma{\rm min}) \approx 23$ s, translating to the cooling length of $\approx 3.2 \times 10^{11}$ cm $\approx 0.1 h < \Delta h= h/3$, consistent with our assumption of the value of $\Delta h$. We find the total radiative power of $L_{\rm j} \approx 1.2 \times 10^{38}$ erg s$^{-1}$. The average energy loss time of all electrons, $E_{\rm e}/L_{\rm j}$ (Table \ref{tab:fits_precj_ssc_maxb}) is then in good agreement with the cooling time scale.

Also, we calculate the Thomson optical depth, $\tau = n_{\rm e} \sigma_{\rm T} R$, where $n_{\rm e}$ is the number density of electrons from $\gamma=1$ to $\gamma_{\rm max}$. We find a low value of $\tau \approx 3.7 \times 10^{-5}$. 

We then calculate the power in both the jet and counterjet. We use eqs.\ (19--21) in \citet{Zdziarski14b}. We assume a homogeneous emitting region (without clumping), pure electron-ion plasma (fully ionized He) without positrons, and a toroidal magnetic field. We first compute the jet power in electrons (excluding their rest energy). We obtain $P_{\rm e} \approx 1.2 \times 10^{38}$ erg s$^{-1}$. Next, we evaluate the total jet power in (cold) iond, as $P_{\rm i} \approx 5.1 \times 10^{38}$ erg s$^{-1}$. The magnetic power is much lower, $P_{B} \approx 3.8 \times 10^{36}$ erg s$^{-1}$. The total jet power is then $\approx 6.3 \times 10^{38}$ erg s$^{-1}$, only slightly lower than that assumed earlier using the model of \citet{ZSSN22}. 

Next, we compute the equipartition parameter, characterizing the degree of equipartition between particles and magnetic field and defined in terms of the energy densities (excluding the rest energy of particles), $\beta_{\rm eq} = u_{\rm e}/(B^2/8\pi)$, where $u_{\rm e} = E_{\rm e}/(\pi R^2 \Delta h)$ is the electron energy density. We find $\beta_{\rm eq} \approx 31$.  Similarly, the magnetization parameter (defined with respect to the rest mass density, see above) is very low, $\sigma \sim 0.001$. This indicates the jet is strongly dominated by the kinetic energy of particles. However, this value of $\sigma$ is lower than that assumed in the model of \citet{ZSSN22}, $\sigma_{\rm min}=0.008$, which is due to the magnetic power in our fit lower than that following from that model. 

In addition, we also compute the mass flow rate in ions in the jet (equation 17 in \citealt{Zdziarski14b}), obtaining $\dot{M}_{\rm j} \approx 4.6 \times 10^{18}$ g s$^{-1}$. We compare this rate to the super-Eddington accretion rate mentioned earlier, $\dot M_{\rm accr}\gtrsim 2\times 10^{19}$\,g\,s$^{-1}$, and the derived jet powers to $\dot M_{\rm accr} c^2 \approx 1.8 \times 10^{40}$ erg s$^{-1}$. We find that the jet is powered by a rather sizeable fraction ($\approx 23$\%) of accretted matter, while only a very small fraction ($\approx 3$\%) of the energy released from accretion ($\dot{M}_{\rm acc} c^2$) is supplied to the particle energy, and an almost negligible fraction is converted into the jet magnetic energy.

Next, we calculate the jet bending angle, $\Phi$, owing to the wind impact. In the limit of $\Phi\ll 1$ rad, we have
\begin{equation}
\Phi\approx \frac{\alpha_{\rm j} \dot M_{\rm w} v_{\rm w}(\Gamma_{\rm j}-1)c}{4\pi \beta_{\rm j}\Gamma_{\rm j} P_{\rm j}},
\label{Phi}
\end{equation}
as given by equation (7) of \citet{Bosch-Ramon16} (see their equation 8 for an expression valid at any $\Phi$). Here $\dot M_{\rm w}$ and $v_{\rm v}$ are the wind mass loss rate and velocity, respectively, and $\Gamma_{\rm j}$ is the jet Lorentz factor. We assume $\dot M_{\rm w}\approx 10^{-5}\msun$ yr$^{-1}$ (\ant) and $v_{\rm w}\approx 1.5\times 10^8$ cm s$^{-1}$ \citep{vanKerkwijk96}. However, since the jet likely undergoes a decelaration over the \g-ray emission region and possibly upstream of it, we use $\Gamma_{\rm j}=2$. This yields a low angle of $\Phi\approx 1\degr$. 

\subsection{Caveats for Model 2}
\label{constraints}

The main problem for this model is the value of the jet wind-induced bending angle, $\Phi \approx 1\degr$, being much lower than the fitted value of $\theta_{\rm j} \approx 41\degr$. If the (unknown) value of $\Gamma_{\rm j}$ is larger, we can increase $\Phi$, but only to $\approx\! 1.8\degr$ for $\Gamma_{\rm j}\gg 1$. Then, the jet half-opening angle in the bending region can be larger, as seen in the hard state \citep{Yang23}. Then, the jet power could be lower depending on the poorly known jet parameters. In particular, if the jet composition consists mostly of e$^\pm$ pairs, $P_{\rm i}$ can be negligible compared to $P_{\rm e}$, which would then increase $\Phi$ by a factor of $\approx$4. Furthermore, values of $\dot M_{\rm w}$ significantly higher than $10^{-5}\msun$ yr$^{-1}$ were obtained by \citet{vanKerkwijk93} and \citet{Ogley01}. In principle, $\Phi \approx 40\degr$ could be reached if $\alpha_{\rm j}$, $\dot M_{\rm w}$ and $\Gamma_{\rm j}$ are higher than the adopted values, and $P_{\rm j}$ is lower.

In addition, if $\Phi < \alpha_{\rm j} = 5\degr$, there would be only a minor modification of the jet propagation due to the thrust of the stellar wind, which would only induce a weak recollimation shock or a sound wave in the jet \citep{Bosch-Ramon16}. However, we argued above that $\Phi$ could be larger, and it appears it could be larger than $\alpha_{\rm j}$, if, in particular, the jet consists of a significant number of e$^\pm$ pairs. 

A major consideration for this model is the required large phase offset, $\varphi_{\rm o}\approx 142\degr\pm 4\degr$. For our assumed masses, the BH velocity around the center of mass is $v_{\rm BH}\approx 6\times 10^7$\,cm\,s$^{-1}$. The average vertical velocity of the jet between the origin and the height of the emission required (via the binary rotation) to reach $\varphi_{\rm o}$ is then $h/(P \varphi_{\rm o}/2\pi)$. Obviously, it is rather low, e.g., $\approx 4.3\times 10^8$\,cm\,s$^{-1} \sim 0.01 c$ for $h=11 a$. This may still possible be if the jet is composed of a slow sheath, formed by a disk outflow, and a fast spine powered by the spin of the rotating BH. 

Another issue is the large fitted polar angle of the orbital precession, $\theta_{\rm j} = 41\degr\pm 2\degr$ in the \g-ray emission region. The jet needs to recollimate above that region in order to satisfy the observational constraint on the semi-opening angle, $\approx\! 5\degr$ \citep{Miller-Jones04, Miller-Jones06}. This could be achieved by reconfinement, for instance one induced by the stellar wind (e.g., \citealt{Yoon15, Yoon16}), and the symmetry and pressure gradients of the environment at the scales of interest. In fact, the large scale jet half-opening angle can become moderate once its interaction with the wind is not significant any more (see, e.g., equation 19 in \citealt{Bosch-Ramon16}). However, due to many scales involved, estimating the jet large-scale properties would require numerical simulations, which are out of scope of this work.

An additional complexity can appear if the magnetic field symmetry axis is inclined with respect to the BH spin axis, e.g., due to interaction with the stellar wind, see, e.g., \citet{James24}. Studying such effects is outside the scope of this paper.

\section{Discussion}
\label{discussion}

After exhaustive research, we are still unable to choose between the two studied models of the observed \g-ray modulation on the orbital period. The blackbody Compton model gives better fits, but it predicts general-relativistic jet precession on the period of 50 years, which is ruled out observationally, see Section \ref{against}. The model with with a jet precessing on the orbital period due to bending by the thrust of the stellar wind fits the data satisfactorily, but the wind is too weak and/or the jet too strong to account for the fitted large bending angle, see Section \ref{constraints}. This phenomenon has been theoretically predicted (e.g., \citealt{Yoon15, Yoon16, Bosch-Ramon16, Barkov22}), but not yet confirmed observationally in any other source. 

An important test possibly allowing us to distinguish between the two models would be either the confirmation or disproval of the presence of orbital modulation with the folded light curve similar to that seen in \g-rays at the mm-wavelength modulation. Such modulation was claimed in a preliminary report of \citep{McCollough23}. If present, it clearly cannot be explained by the Compton anisotropy effect present in our Model 1, but it can be due to the Doppler boosting present in our Model 2. However, we note that modulation at mm wavelengths could also be due to free-free absorption in the stellar wind, and the orbital phase of the maximum absorption can be shifted due to the jet misalignment with respect to the orbital axis.

We consider then the origin of the structures observed by \citet{Mioduszewski01} and \citet{Miller-Jones04}, which, if caused by precession, would have much longer periods than the binary one, $\gtrsim$60\,d and $5.5\pm 0.5$\,d, respectively. We speculate that these long periods are the result of perturbations in the jet on large scales triggered by instability growth as the jet is being perturbed at its base by the effect of the wind and the orbital motion. This would give rise to a distinct pattern not synchronized with the orbital cycle but rather representing a subharmonic of it. Specifically, if the jet bending is excited with a wavelength $N$ times the orbital period multiplied by the jet velocity, the resulting spatial scale would be $\sim 10^3 N a$.

A consequence of Model 2 is that the jet (on average) and the binary should have the same inclination. The binary inclination determined by \ant is $i=29\fdg5\pm 1\fdg2$. The jet inclination determined by \citet{Miller-Jones04} based on the proper motion of the approaching and receding components assuming they are intrinsically symmetric \citep{MR94} and $D=9$\,kpc is $i_{\rm j}\approx 26\degr\pm 6\degr$, i.e., consistent with the above value of $i$. On the other hand, the jet inclination based on fitting the precessing jet model of \citet{HJ81} to the curvature observed in the VLBA images of Cyg X-3 is $i\approx 10\fdg4\pm 3\fdg8$ \citep{Miller-Jones04} and $i\lesssim 14\degr$ \citep{Mioduszewski01}. Those two estimates are incompatible with the inclination of \ant, which would then favor Model 1.

In Model 2, we considered the \g-ray emission region at a relatively large distance from the origin, $h\approx 10 a$, which allowed us to obtain the synchrotron emission strong enough to account for the observed average 225\,GHz flux. This distance is much larger than that expected for a recollimation shock from the jet-wind interactions, which is at $h\sim (1$--$2) a$ \citep{Yoon16, Bosch-Ramon16, Barkov22}. Then, the dissipation at $h\sim 10 a$ would need to be explained by another process. Alternatively, we can relax the requirement of reproducing the 225\,GHz flux by assuming that emission originates in a different part of the jet. Then, the SSC emission alone in that model is independent of $h$, and we could have $h\sim (1$--$2) a$. 

\section{Conclusions}

Our main conclusions are as follows.

We have used a \g-ray bright data set from the LAT observations of Cyg X-3 greatly enlarged with respect to the previous study of \aaz. We have found that the \g-rays are modulated at the period that is compatible with the orbital period based on the X-ray data at the accuracy of $<0.02$\,s (accounting for the secular period increase).

We have studied possible models of the orbital modulation. We have significantly improved the model based on anisotropy of Compton scattering, by including KN effects and the presence of a magnetic field. The model fits well both the modulation profile and the average spectrum, and requires that the jet is misaligned with respect to the orbital axis by $\sim\! 35\degr$. If the jet direction follows the BH spin axis, both will undergo the geodetic precession with a period of $\sim$50\,yrs. Its presence is ruled out by both the \g-ray and radio data. If this is the correct model, the jet misalignment is not related to the BH spin axis.

Therefore, we have proposed an alternative model using the jet bending due to the thrust of the stellar wind from the donor. The jet is aligned with the orbital axis on average, but it is bent to outside of the orbit, causing its precession at the orbital period. The Doppler boosting of both the synchrotron and SSC emission is then orbitally modulated. The modulated SSC emission fits well the \g-ray data. The modulation of the SSC in this model does not depend on the distance of the emission region from the center. However, the synchrotron flux (at the SSC flux determined by the \g-ray data) depends on it. If we postulate that the observed average mm flux is reproduced by this model, this distance is about 10 orbital separations. The mm emission can still be from a different jet region, and then the \g-ray emission distance can be lower. However, a major problem for this model is that the theoretically predicted bending angle at plausible wind and jet parameters is much lower than that fitted to the data. 

In both models, the jet velocity is low, $\beta_{\rm j}\sim 0.5$, and the jet is weakly magnetized. The fitted field strengths are well below that predicted by models of the magnetically arrested disk and the magnetically launched jet. 

\section*{Acknowledgements}
We thank I.~Antokhin, E.~Egron, M.~Liska and M. McCollough for discussions. We acknowledge support from the Polish National Science Center grants 2019/35/B/ST9/03944, 2023/48/Q/ST9/00138, and from the Copernicus Academy grant CBMK/01/24. The work of AD is supported by the South African Department of Science and Innovation and the National Research Foundation through the South African Gamma-Ray Astronomy Programme (SA-GAMMA).

\bibliographystyle{aasjournal}
\bibliography{../../../allbib.bib} 

\end{document}